\documentclass[11pt]{elsarticle}
\usepackage{epsfig,graphics,amssymb,amsmath,graphicx,indentfirst,subfig,float,appendix,multirow,pifont,color,soul,esint,tabularx}
\usepackage{booktabs}
\usepackage{threeparttable}
\usepackage{graphicx}
\usepackage{placeins}
\usepackage[table,xcdraw]{xcolor}
\usepackage{lineno}
\newcommand{\bl}[1]{\boldsymbol{#1}}

\begin{document}
\begin{frontmatter}

\title{Investigating the effect of turbulence on hemolysis through cell-resolved fluid-structure interaction simulations}

\author[cornell]{Grant Rydquist}
\author[cornell]{Mahdi Esmaily$^*$}
\address[cornell]{Sibley School of Mechanical and Aerospace Engineering, Cornell University, Ithaca, NY 14850, USA}

\date{\today} 
\begin{abstract}
Existing hemolysis algorithms are often constructed for laminar flows that expose red blood cells to a constant rate of shear. It remains an open question whether such models are applicable to turbulent flows, where there is a significant variation in shear rate along cell trajectories. To evaluate the effect of turbulence on hemolysis, we perform cell-resolved simulations of red blood cells in turbulent channel flow at $Re_\tau=180$ and 360 and compare them against the results obtained from laminar flow simulations at an equivalent wall shear stress. This comparison shows that, while the laminar flow generally induces greater stretch in the cell in a time-averaged sense, cells experience an overall larger deformation in turbulence. This difference is attributed to extreme events in turbulence that occasionally create bursts of high shear conditions, which, consequently, induce a large deformation in the cells. Associating damage with the most extreme deformation regimes, we observe that, in the worst case, the turbulent flow can produce deformation in the cell that is higher than the absolute maximum value in the analogous laminar case approximately 14\% of the time. Additionally, the $Re_\tau=180$ universally induced greater deformation in the cells than the $Re_\tau = 360$ case, suggesting that increasing the range of scales in the flow does not necessarily yield greater deformation when all other parameters are kept constant. A strong direct correlation ($R>0.8$) between shear rate and deformation metrics was observed in turbulence. The correlation against $Q$-criterion is inverse and weaker ($R\approx -0.26$), but once the shear contribution is subtracted, it improves in terms of areal dilatation ($R\approx -0.6$). 

% END CONTENT ABS------------------------------------------
%\noindent
%\textit{\textbf{Keywords: }%
%Stokes equation; Complex-valued solver; Cardiovascular flow;} \\ %% <-- Keywords HERE!
\end{abstract}
\end{frontmatter}

% %%%%%%%%%%%%%%%%%%%%%%%%%%%%%%%%%%%%%%%%%%%%%%%%%%%%%%%%%%
% %%%%%%%%%%%%%%%%%%%%%%%%%%%%%%%%%%%%%%%%%%%%%%%%%%%%%%%%%%
% BODY OF THE DOCUMENT
% %%%%%%%%%%%%%%%%%%%%%%%%%%%%%%%%%%%%%%%%%%%%%%%%%%%%%%%%%%
% %%%%%%%%%%%%%%%%%%%%%%%%%%%%%%%%%%%%%%%%%%%%%%%%%%%%%%%%%%

% --------------------
\section{Introduction} \label{sec:introduction}

% -- What's been done/why am i doing this
The damage induced by implanted cardiovascular devices and surgical treatments on red blood cells (RBCs), known as hemolysis, plays a major role in the design process for the risk that it poses to patient health. To mitigate such risks, various computational fluid dynamics (CFD) based techniques have been introduced to predict the amount of damage a certain design will inflict on RBCs \cite{giersiepen_estimation_1990, sharp_scaling_1998, garon_fast_2004, arvand_validated_2005, arora_tensor-based_2004, nikfar_prediction_2020, sohrabi_cellular_2017, ezzeldin_strain-based_2015, arwatz_viscoelastic_2013}. From a mechanistic point of view, hemolysis is a complex process, spanning from small pores forming in the membrane, leading to permeabilization of the membrane \cite{hallow_shear-induced_2008, schlicher_mechanism_2006}, to large-scale fragmentation of the cells \cite{sutera_deformation_1975}, and it depends on factors such as the history of the cells as well as individual cell properties. As a result, hemolysis prediction remains an active area of research. 

One particular area of research where disagreement persists is the effect turbulent flows have on RBCs. Some pioneering experiments have shown that turbulent flows tend to inflict more damage than laminar flows on RBCs under similar conditions \cite{kameneva_effects_2004}. More specifically, these early experiments are performed by running RBCs through a pipe at different Reynolds number and measuring the level of plasma-free hemoglobin after a certain period of time. The key point in these experiments is that the Reynolds number is altered by adjusting the fluid viscosity through the addition of Dextran-40. This way, the pressure drop across the pipe is kept the same between turbulent and laminar experiments, thereby ensuring that, on average, RBCs experience the same shear stress as they pass through the pipe. While these early experiments indicate greater hemolysis in the presence of turbulence, more recent experiments that are currently underway have been inconclusive in this regard (personal communications with collaborators attempting to replicate the results of Kameneva et al). Therefore, more experimental observation is needed to establish the effect of turbulence on hemolysis. 

Computationally, many existing hemolysis algorithms are empirical and fitted based on data from laminar flows. Given that the cells experience fundamentally different conditions in a laminar and turbulent flow, these laminar-based algorithms can be unreliable if applied incorrectly to turbulent flows. Some algorithms have been introduced to characterize hemolysis in turbulence \cite{tamagawa_prediction_1996, goubergrits_turbulence_2016, ozturk_approach_2016, yen_effect_2014}; however, these algorithms still lack consensus and lack a mechanistic approach to systematically model the underlying physics causing damage. Many of these algorithms use the Reynolds stress to characterize the amount of damage to the cells. However, as pointed out by Quinlan \cite{quinlan_mechanical_2014}, these stresses are macroscopic averages that do not take into account the microscopic details of the flow and do not represent a physical stress experienced by the cells.

Looking in greater detail at the cells' actual responses to the flows could provide insights into how best to incorporate the effects of turbulence into these types of algorithms, or as a means to evaluate hemolysis in itself in a way that is agnostic to whether the flow is laminar or turbulent. The purpose of the current work is to investigate the question of how RBCs respond to turbulence from a cell-resolved point of view and compare that response to RBCs deforming under similar conditions in laminar flow to gain a mechanistic understanding of the differences between these flow types at a cellular level. Since these mechanisms are not well understood, cell-resolved simulations could provide insights by showing the full profile and history of the deformation of the cell. 

There have been attempts to characterize the effect of turbulent flows on RBCs in previous studies \cite{dooley_effect_2009, m_faghih_characterization_2018}, which provide some useful order-of-magnitude arguments on the importance of some stresses. However, these studies relied on simplified representations of the cell or turbulent flow. For example, the first of these is performed in 2D using a circular membrane and idealized, circular eddies, and the second of these studies uses estimates of the scalar stress in turbulent flow to obtain estimates of the tension in the RBC membrane. In the latter case, direct simulations are not performed. Both of these studies lack a direct representation of the chaotic motion of turbulence, instead using an artificially created flow, and a well-resolved representation of the cell membrane. To the author's best knowledge, the present study is the most faithful reconstruction of a RBC's response to turbulence in macroscale flows.

% -- Introduction of how I plan to do this (compare proposed methods, talk about solver). moved to methods
% -- Outline of papaer
The paper is outlined as follows: first, the methods for the study are introduced in Section \ref{sec:method}.  Largely, the computational methods in this study are based on previous works \cite{rydquist_cell-resolved_2022}. The flow conditions are modeled after the experimental work of Kameneva et al \cite{kameneva_effects_2004}. Next, results are presented in Section \ref{sec:results}, with comparisons drawn between the laminar and turbulent cases. Results are presented primarily in terms of probability density functions showing the likelihood of the maximum deformation a cell will experience at any given point in time. In Section \ref{sec:discussion}, the results are discussed and analyzed. Future work is discussed, and finally, conclusions are presented in Section \ref{sec:conclusions}.

% --------------------
\section{Methods}\label{sec:method}
\begin{figure}[H]
  \centering
  \includegraphics[width=0.8\textwidth]{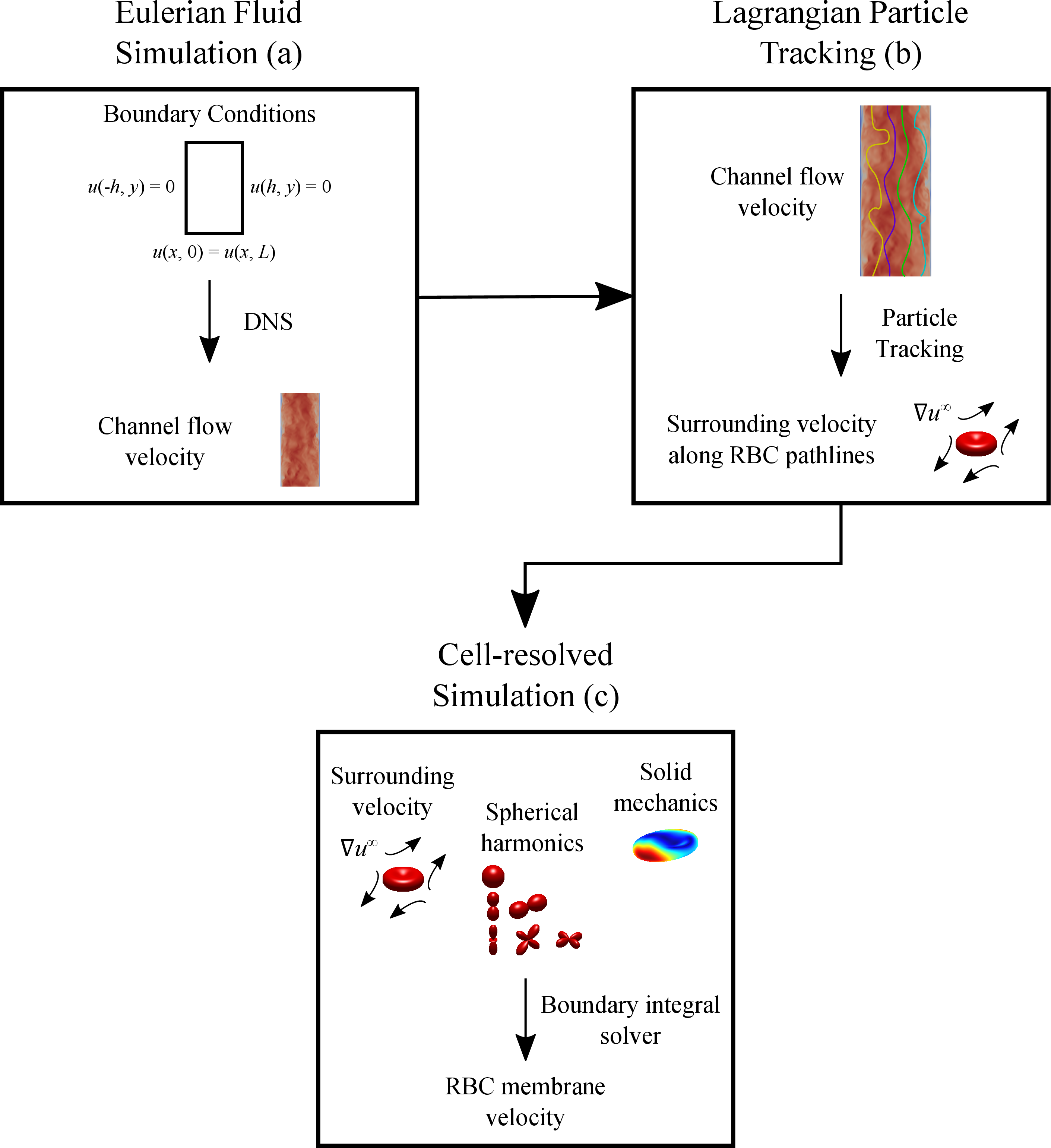}
  \caption{The steps involved in the numerical framework to simulate the response of individual RBCs to a macroscale flow. First, direct numerical simulation is used to resolve the fluid dynamics in the macroscale flow of interest (a). Here, a turbulent channel flow is simulated. Next, particles are randomly seeded in the channel, and the velocity gradient is tracked as the particles move through the flow (b). These velocity gradients are used as an input into a boundary integral solver to solve for the velocity of the individual RBCs (c).}
  \label{fig:schematic}
\end{figure}

To facilitate the comparison of these flows, we will use a multiscale computational framework for resolving individual RBC behavior in macroscale flows outlined in a previous work \cite{rydquist_cell-resolved_2022}. In short, this simulation relies on three separate solvers: first to simulate the large-scale fluid dynamics, second to model trajectories of the cells in the flow, and third to solve for the response of the RBCs to the flow in their immediate vicinity. These individual steps are outlined in Fig. \ref{fig:schematic}: (a) the fluid flow of interest is simulated using direct numerical simulation; (b) Lagrangian particles are tracked through the flow, and the velocity gradient is tracked; and (c) the velocity gradient, which represents the flow field surrounding RBCs, is employed as an input to simulate RBCs dynamics using a boundary integral solver. In the following three sections, these three steps are explained in detail. 

% - - - - - - - - - - -
\subsection{DNS of Channel Flow}
The general setup for comparing the effect of laminar and turbulent flows on RBCs follows the methodology of Kameneva et al \cite{kameneva_effects_2004}, who were among the first to experimentally study this effect quantitatively. The key element in their methodology in performing this comparison was to keep the wall shear stress the same between laminar and turbulent flows. This way, the effect of turbulence on hemolysis can be studied independently of the shear stress. 

Even though the early experiments mentioned above are performed in a pipe, we consider a channel flow configuration in the present study. This entails two infinite parallel plates with periodic boundary conditions imposed on the streamwise and spanwise directions and no-slip boundary conditions imposed on the walls. This canonical configuration was selected for its historical significance as it is the simplest wall-bounded turbulent flow that one can generate. Since the smallest scales of motion in the fluid are likely to be important to the deformation of the RBCs, we employ direct numerical simulation (DNS). This step corresponds to Fig \ref{fig:schematic}a. This solver is described in \cite{esmaily2020benchmark,esmaily2018scalable}, and follows the methodology of the seminal work by Kim et al \cite{kim_turbulence_1987}. 

\begin{figure}[htbp]
  \centering
  \includegraphics[height=0.5\textwidth]{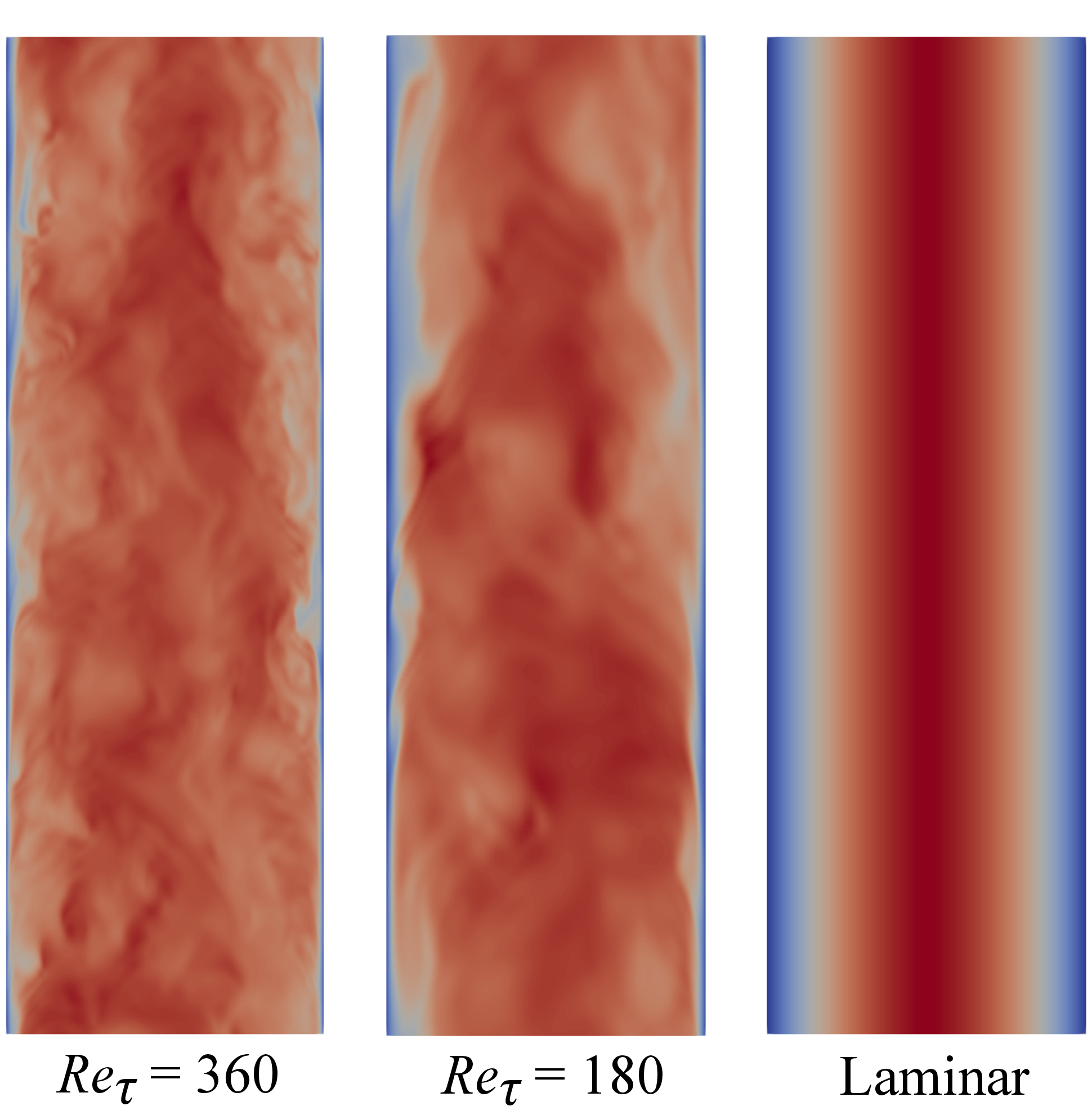}
  \caption{Snapshots the streamwise velocity magnitude of the turbulent and laminar cases used in this study. The turbulent cases are for $Re_\tau = 180$ and $360$. Note that the laminar case displays a parabolic velocity profile.}
  \label{fig:turbchan}
\end{figure}

For the laminar case, a parabolic flow profile was utilized. Fluid simulations were not run for this case as laminar channel flow has a known analytical form of $u(y) = \left[1-(y/h)^2\right]\tau_wh/2\mu$. Here, $h$ is the channel half-width and $\tau_w$ is the wall shear stress. Two turbulent cases are run: one at friction Reynolds number $Re_\tau=180$ and another at $Re_\tau=360$. Snapshots of the laminar and turbulent cases are displayed in Fig. \ref{fig:turbchan}. Simulations are compared at wall shear stresses of 10, 20, 30, and 40 Pa.

Numerical parameters for the turbulence simulations are listed in Table \ref{tab:params}. Here, $x$, $y$, and $z$ are the streamwise, wall-normal, and spanwise directions, respectively. The grid is stretched in the wall-normal direction for increased resolution near the wall, and the range of grid spacing $\Delta y$ in this direction is provided. In this table, $\Delta t$ is the time step size, $N$ is the number of grid points in a direction, and $L$ is the length of the channel non-dimensionalized by the channel half-width $h$. Parameters reported in wall units are distinguished by a plus superscript$^+$, e.g., $x^+=xu_\tau/\nu$ and $t^+= t\tau_w/\mu$. Here, $u_\tau=\sqrt{\tau_w/\rho}$ is the friction velocity, $\rho$ is the density, and $\nu$ is the kinematic viscosity. 
In total, we run these simulations to ensure they are statistically converged. 
After that, they are continued for 1,000,000 and 100,000 time steps at $R_\tau=180$ and 360, respectively, to produce a sufficiently long temporal window from which velocity gradient statistics can be collected. 

\begin{table}[H]
\centering
\caption{Numerical parameters used in the turbulent simulations.}
\label{tab:params}
\begin{tabular}{l|l|l}
                          & {$Re_\tau = 180$}         & {$Re_\tau = 360$}  \\ \hline
$L_x\times L_y\times L_z$ & $2\pi\times2\times\pi$    & $2\pi\times2\times\pi$    \\
$N_x\times N_y\times N_z$ & $192\times 128\times 128$ & $384\times 256\times 256$ \\
$\Delta t^+$       & $4.86\times 10^{-2}$       & $5.18\times 10^{-2}$      \\
Range of $\Delta y^+$           &  $0.418 - 5.74$   &    $0.407-5.83$      \\
$\Delta x^+$           &  $5.89$   &    $5.89$      \\
$\Delta z^+$           &  $4.42$   &    $4.42$                     
\end{tabular}
\end{table}

There are some differences between the current work and the experiments of Kameneva et al \cite{kameneva_effects_2004}, which is one of the seminal papers detailing these differences in hemolysis between laminar and turbulent flows under similar conditions. First, as mentioned earlier, that work examined pipe flow, whereas the current work is examining channel flow. We do not anticipate this change in geometry to significantly affect our conclusions as the comparison between turbulent and laminar flow cases is performed on a relative basis. The bulk Reynolds number based on average streamwise velocity and full channel height was 5,680 and 12,470 for $Re_{\tau}=180$ and 360 cases, respectively. This range is comparable to, albeit moderately greater than, the range of bulk Reynolds numbers (2,230-5,100) that were considered in the experiments by Kameneva. Secondly, the wall shear stresses that are used in this work are significantly lower than those used in the experiments by Kameneva et al (100-400 Pa), and in fact, are lower than the typical threshold stresses under which hemolysis typically occurs (above roughly 100-200 Pa) \cite{yu_review_2017}. The RBC solver had issues with long-term stability above a wall shear stress of approximately 40 Pa, after which a small but significant portion of the simulations failed. In this range, the RBCs underwent extreme deformations and took on complex shapes, which in reality, may indicate cell rupture. Provided that our computational framework does not model rupture, and that attempting to resolve the behavior of the RBCs in the regime where they hemolyze would not faithfully reproduce their behavior, we consider a lower shear stress to avoid such conditions. Thus, 40 Pa marks the largest value of wall shear stress used in this study. It is anticipated that differences between flow types will be visible in simulations below the stresses in which the cells begin to hemolyze.

% - - - - - - - - - - -
\subsection{Lagrangian Particle Tracking}
The channel flow simulations are followed by Lagrangian particle tracking to obtain the flow conditions experienced by the cells (corresponding to Fig \ref{fig:schematic}b). In the below simulations, 200 mass-less tracers are randomly seeded in the DNS channel flow and tracked, using the same time step as the fluid simulation. The particles are located each time step using an algorithm described in a previous work \cite{rydquist_optimal_2020}. The velocity gradient at the location of the tracers is recorded, and this is used as an input for the cell-resolved simulations. Velocity gradient information is outputted every 10 time steps of the DNS simulations. In total, the velocity gradient is recorded for approximately 500,000 and 50,000 wall unit time at $Re_\tau=180$ and $360$, respectively, although this entire range is not used for the cell-resolved simulations. By conservation of mass, the tracers should remain uniformly distributed throughout the flow over time. In order to ensure this, 10,000 particle tracers were tracked over multiple time steps for the $Re_\tau=180$ simulation, and their distance from the wall was recorded. The channel was split into 50 evenly spaced bins, and the average number of particles in each bin, normalized by the mean number of particles per box, is reported in Fig. \ref{fig:bins} below. The distribution is fairly uniform, particularly when compared against similar distributions where particles migrate towards the wall due to the effects of turbulence known as turbophoresis \cite{reeks1983transport, esmaily2020benchmark, rydquist_optimal_2020}.

\begin{figure}[htbp]
  \centering
  \includegraphics[height=0.33\textwidth]{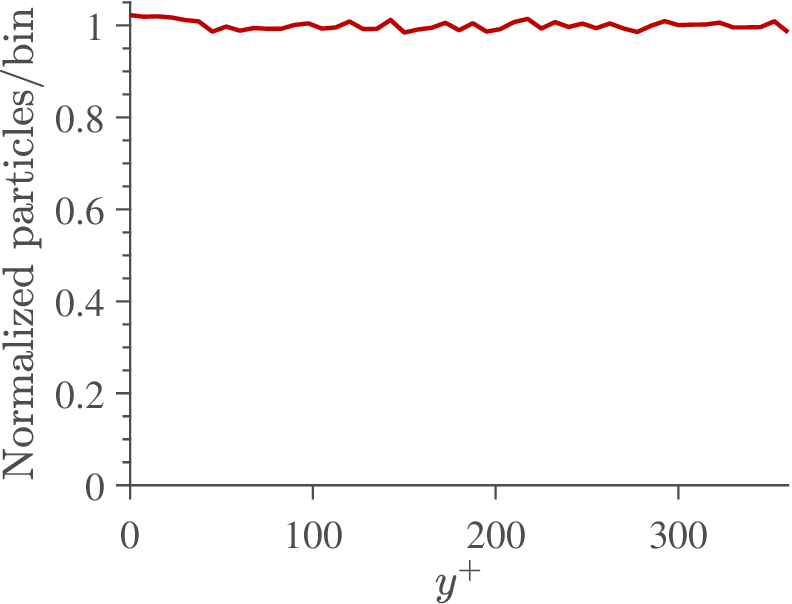}
  \caption{The average distribution of particles as a function of $y^+$ for the $Re=180$ case. The channel is split into 50 equally spaced bins, and each marker represents the number of particles in each bin normalized by the mean number of particles per bin.}
  \label{fig:bins}
\end{figure}

\begin{figure}[H]
  \centering
  \includegraphics[height=0.4\textwidth]{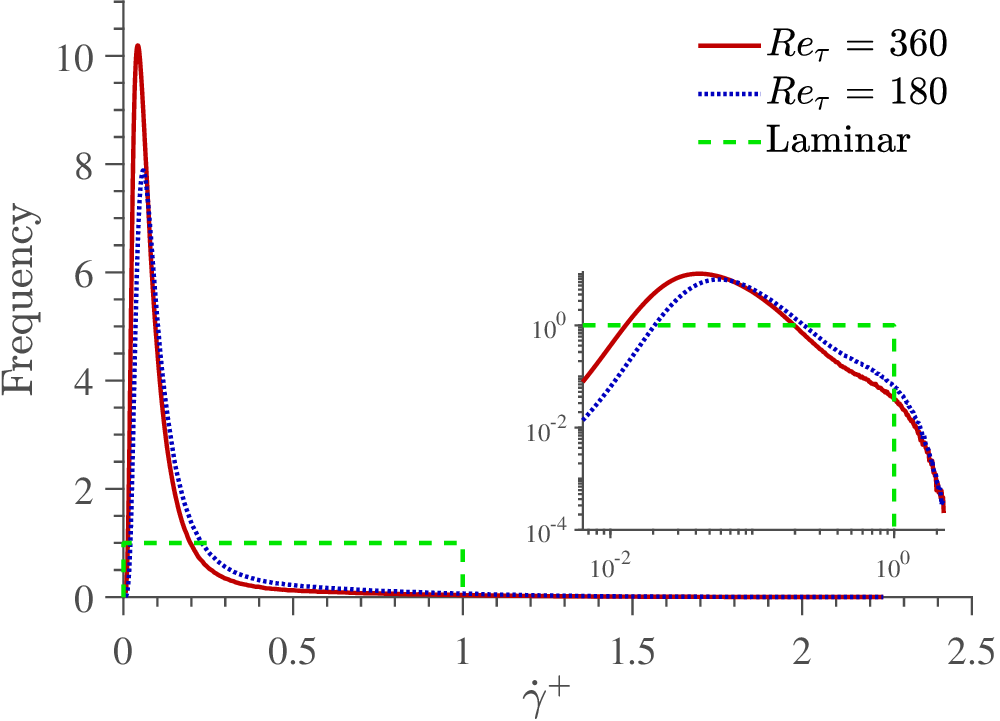}
  \caption{A probability density function of the velocity gradient inputs into the boundary integral solver, normalized by the wall shear stress. The turbulent distributions are indicative of the viscous stress, and have an expected value that is much less than the laminar distribution. The inset figure shows the same plots in a log-log scale, where the velocity gradients rapidly drop above a value of $\dot\gamma^+ = 1$.}
  \label{fig:inputs}
\end{figure}

Figure \ref{fig:inputs} shows a measure of the inputs into the RBC simulations gathered from this particle tracking: a probability density function (PDF) of the norm of the velocity gradient tensor, given as $\dot \gamma = \sqrt{\nabla_iu_j\nabla_iu_j}$, normalized by the characteristic strain rate in the flow, $\tau_w/\mu$ such that $\dot\gamma^+=\dot\gamma\mu/\tau_w$. These PDFs are provided for the three cases under investigation: two turbulent cases at friction Reynolds numbers $Re_\tau = 180$ and $360$ and the laminar case. These PDFs are universal across all values of the wall shear stress. One interesting feature of these PDFs is the fact that the norm of these velocity gradients is concentrated at significantly smaller values in the turbulent cases, with this being more pronounced in the higher Reynolds number case. This is because these values are dominated by the viscous stress, as they are collected via a temporal ensemble average where the time-average value of $\nabla_iu_j$ diminishes quickly outside the viscous sublayer. Some variation is also introduced via the Reynolds stress. Additionally, a log-log scale has been provided, which makes the tails of the distribution more apparent. While there are some locations where the magnitude of the shear rate is greater than at the wall in the turbulent case, the frequency drops off particularly rapidly above $\dot \gamma^+ =1$.

% - - - - - - - - - - -
\subsection{Cell-resolved Simulations}
The velocity gradient along the RBC's trajectory from the macroscale flow is imposed as the far-field velocity to simulate the RBC deformation. This step is completed using a boundary integral method coupled to a structural solver (corresponding to Fig \ref{fig:schematic}c). Because the time steps did not line up between the recorded velocity gradient files and cell-resolved simulations, quadratic interpolation was used when obtaining the velocity gradient for the RBC simulations. Although the solver is capable of handling cell-cell interactions and periodic boundary conditions, the simulations are run for isolated cells in an unbounded domain, as an upcoming work has shown that the first-order effects of cell-cell interactions can be captured by properly adjusting the fluid viscosity to reflect the suspended RBC phase \cite{rydquist2023analysis}. The results of this work have shown that, as long as many potential orientations of the RBC are simulated and averaged over and the viscosity is taken to be the effective viscosity of the fluid, these single-cell simulations well-approximate the deformation of RBCs in multiple-cell, periodic simulations \cite{rydquist2023analysis}. As such, when seeding the RBCs into the flow, the orientation is taken to be random with respect to the shearing direction. The hematocrit of the experiments of Kameneva et al was 24\%, corresponding well to the simulations in this upcoming work \cite{rydquist2023analysis}. This process provides a well-resolved picture of the RBCs' response to the flow, while also providing a complete history of the cells' deformation, which could be useful for processes like pore formation, leakage, and full-scale rupture.

The RBC solver is described in a previous work, with details deferred to that paper \cite{rydquist_cell-resolved_2022}. In brief, the solver represents the RBC membrane as a two-dimensional continuum in three-dimensional space, an assumption made due to the thinness of the RBC membrane relative to its other characteristic lengths. This, along with a low-Reynolds number assumption in the direct neighborhood of the cell, facilitates the use of a boundary integral method to solve for the motion of the RBC, given internal membrane stresses and the surrounding fluid velocity. This method is relatively inexpensive, since only the motion of the RBC membrane needs to be resolved, and not that of the surrounding fluid. The shape of the RBC is represented using spherical harmonic basis functions, which allow fast and accurate evaluation of some of the integrals required in the boundary integral method.

Simulations are run at a spherical harmonic order of $p=12$. The RBC simulations are run using a time step equal to $\Delta t^+ = 0.01$. To remain consistent across cases, the wall shear stress is used to non-dimensionalize the problem, which enters into the Capillary number, a measure of the relative strength of the flow to the cell's resistance to deformation, as $Ca = \tau_w a/G$. Here, $a = 2.82$ $\mu$m is the characteristic radius of the cell, defined as the radius of a sphere with equivalent volume, and $G=2.5$ $\mu$N/m is the cell's shear elastic modulus.

The viscosity ratio used for this work is $\lambda = \mu_{in}/\mu_{out} =1$, where $\mu_{in}$ is the interior cytoplasm viscosity and $\mu_{out}$ is the effective viscosity of the surrounding medium. Commonly, a cytoplasm viscosity $\mu_{in}$ of approximately 6 cP is used for the viscosity of the interior cytoplasm viscosity, although there is little experimental work and some uncertainty on this value \cite{zhao_spectral_2010, pozrikidis_numerical_2003, sohrabi_cellular_2017, tomaiuolo_biomechanical_2014}. The effective viscosity of the solution given in experiments by Kameneva et al is 6.3 $\pm$ 0.1 cP for the laminar case and $2.0 \pm 0.1$ cP for the turbulent case. Thus, this value of the viscosity ratio corresponds approximately to the laminar case in the experiments. Note that since the turbulent experimental cases utilized a lower effective viscosity of the surrounding fluid, this would correspond to a greater value of the viscosity ratio. However, the decision to use the same viscosity ratio for both the laminar and turbulent cases was made deliberately to facilitate a more direct comparison between the results and eliminate this parameter as a possible confounding variable.

Table \ref{tab:rbcpar} shows the ranges of the non-dimensional parameters used for the RBC simulations, along with their definitions. Also included in this table is $Ca_\eta = \mu  a/Gt_\eta$, the Capillary number defined with respect to the Kolmogorov timescale, which provides a second measure of the RBCs' properties relative to those of the flow. This value can be derived directly from $Ca$ by noting that $t_\eta = \sqrt{\nu/\varepsilon}$, where $\nu$ is the kinematic viscosity and $\varepsilon$ is the energy dissipation rate per unit mass. Further, the average value of $\varepsilon$ can be written as $\frac{\mathrm d P}{\mathrm d x}\frac{U_b}{\rho}$, where $\frac{\mathrm d P}{\mathrm d x}$ is the average pressure drop, $U_b$ is the bulk velocity, and $\rho$ is the density. Finally, the average pressure drop must be balanced by the average wall shear stress as $\frac{\mathrm d P}{\mathrm d x} = -\tau_w/h$, such that $Ca_\eta$ can be written as, after some arithmetic, $Ca_\eta = Ca\sqrt{\mu U_b/\tau_w h}$. As such, a value is provided for the laminar flow as well.

\begin{table}[H]
\centering
\caption{The ranges of non-dimensional parameters used in the cell-resolved simulations. The depicted range corresponds to varying the wall shear stress from 10 to 40 Pa. }
\label{tab:rbcpar}
\begin{tabular}{l|l|l|l}
          & $Re_\tau = 360$ & $Re_\tau = 180$ & Laminar       \\ \hline
$Ca=\tau_w a/G$      & $11.3-45.1$   & $11.3-45.1$   & $11.3-45.1$ \\
$Ca_\eta =\mu a/Gt_\eta$  &        $2.22-8.78$           &      $2.96-11.8$       &     $5.77-23.1$          \\
$C = E_d/G$       & $451.2-1,810$   & $451.2-1,810$   & $451.2-1,810$ \\
$\lambda =  \mu_{in}/\mu_{out}$ & 1               & 1               & 1            
\end{tabular}
\end{table}

Note that there is some disagreement about the value of the RBC shear elastic modulus, which is an essential component used to nondimensionalize this problem. For example, the 2014 review by Tomaiuolo, which compiled geometric and mechanical properties of healthy RBCs, listed a range of the shear elastic modulus of 5.5 $\pm$ 3.3 $\mu$N/m \cite{tomaiuolo_biomechanical_2014}. Selection of the value of this parameter can thus have a significant impact on the interpretation of the results. For example, $Ca =10$ could correspond to a value of $\tau_w= 7.80-31.21$ Pa. The current value of 2.5 $\mu$N/m was selected based on the analysis of Dimitrakopoulos on the Skalak constitutive model used in this work \cite{dimitrakopoulos_analysis_2012}, who suggests that differences in the measurements of this parameter can be explained by differences in the particular constitutive model being used to fit the experimental measurements (for example, fitting measurements with a neo-Hookean constitutive model might not yield the same value of $G$ as measurements fit with the Skalak model). This value is on the lower end of the above range, however. Other works have used a larger value of this parameter and obtained results that match well with experiments and computational benchmarks \cite{sinha_dynamics_2015, zhao_spectral_2010, fedosov_multiscale_2010}.

The main parameter of interest to be reported below in the results section is $\max(\lambda_1/\lambda_2)$, where $\lambda_{1,2}$ are the principal stretches on the surface of the cell, thus providing a measure of local stretch. Here, the $\max$ is taken over the entire surface of an individual cell at a given point in time. This parameter is represented by 
\begin{equation}
S\equiv\max(\lambda_1/\lambda_2).
\end{equation}
Additionally, the parameter $\max(\lambda_1\lambda_2)$ is investigated, representing the maximum local area dilatation ratio on the surface of the cell. The maximum of these parameters is of particular interest because it is assumed that hemolytic events will occur at these locations.

It should be noted that the values of $\max(\lambda_1\lambda_2)$ are likely elevated compared to what would be observed in reality. This is because the value of the dilatation ratio $C$ used in these simulations is smaller than in actual RBCs. $C$ is the ratio of the dilatation modulus $E_d$ to the shear elastic modulus $G$. This parameter represents the cell's resistance to local area changes. In reality, RBCs strongly resist changes in area, but the large value of this parameter causes the underlying equations to become particularly stiff. As a result, a lower value of this parameter is typically used in computational studies \cite{yazdani_phase_2011, zhao_spectral_2010, rydquist_cell-resolved_2022, pozrikidis_numerical_2003}. However, it has been shown that using a smaller value of this parameter, even decreasing its value by more than two orders of magnitude, produces relatively well-converged behavior of large-scale RBC dynamics \cite{rydquist_cell-resolved_2022, sinha_dynamics_2015, barthes-biesel_effect_2002, pozrikidis_numerical_2003}. Thus, it is assumed that comparisons of $\max(\lambda_1\lambda_2)$ will be valid across simulations at the same wall shear stress. This value is also salient, as it is often used as a parameter of interest in cell-scale hemolysis \cite{xu_cell-scale_2023, nikfar_multiscale_2020}. Additionally, as the wall shear stress is increased, the stiffness of the problem becomes less severe. This is because the relevant timescales of the flow approach the timescale associated with the area dilatation. Thus, the area dilatation ratio is scaled linearly with the Capillary number. At a Capillary number of $Ca=1$, a value of $C=40$ is used. This is comparable to previous works \cite{zhao_spectral_2010, pozrikidis_numerical_2003} when accounting for the nondimensionalization of this parameter. At the higher end of the values of $Ca$ used, the value of $C$ begins to approach, but does not yet reach, the physiological value of approximately $10^5$. A second practical consequence of this scaling, however, is that $\max(\lambda_1\lambda_2)$ also displays relatively less change as the wall shear stress is increased than would be expected in reality, as the resistance to area change is higher in flows with greater wall shear stress. In order to make some accommodation for this, the value
\begin{equation}
A\equiv \frac{C}{C_0}\left[\max(\lambda_1\lambda_2)-1\right]
\end{equation}
is reported below, where $C_0=10^5$ is the physiological value of the dilatation ratio, and $C$ is the value actually used in the simulations.

In summary, simulations are performed at three Reynolds numbers (laminar, and turbulence at $Re_\tau=180$ and 360) and four wall shear stress ($\tau_w=10$, 20, 30, and 40 Pa) for a total of 12 flow configurations. In what follows, we describe the outcome of these calculations. 

% --------------------
\section{Results} \label{sec:results}
Our computations are performed over time for many cells. To process the results, multiple types of averages are utilized. First, an average can be performed across all cells, denoted $\langle \bullet\rangle_c$. This operation produces a time-varying average of the parameters of interest. Additionally, an average across a single cell over all time, denoted $\langle \bullet\rangle_t$, could be performed, which would produce a single value for each cell. This operation is not used explicitly in these results. Finally, an average over all cells and all time can be performed, denoted simply $\langle \bullet\rangle \equiv \langle \langle \bullet \rangle_c \rangle_t$, which produces a single value for the entire simulation.

\begin{figure}[H]
    \centering
    \subfloat[ \label{sub:indivs360}]  {\includegraphics[scale=0.4]{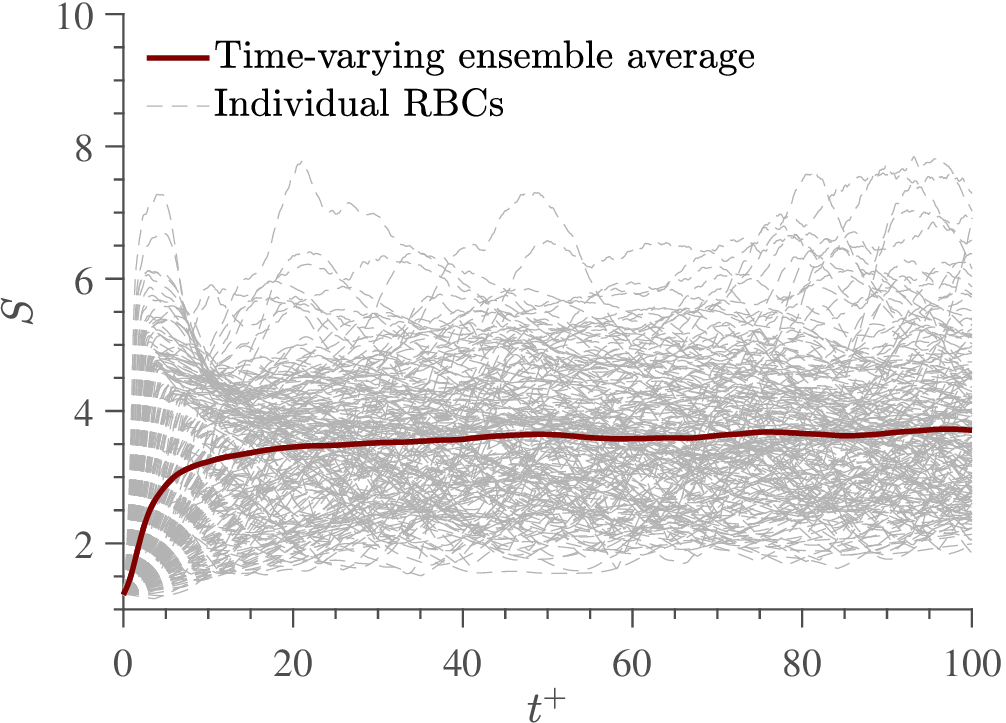}}\quad%
    \subfloat[ \label{sub:indivs360A}] {\includegraphics[scale=0.4]{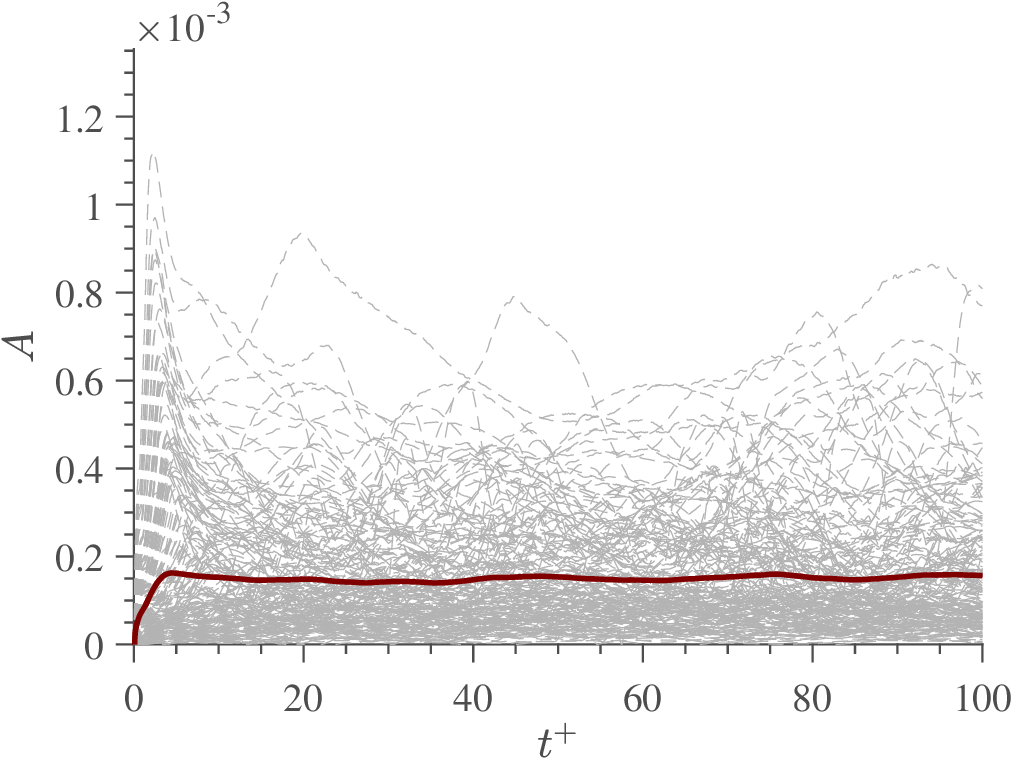}}\quad\\%
    \subfloat[ \label{sub:indivs180}]   {\includegraphics[scale=0.4]{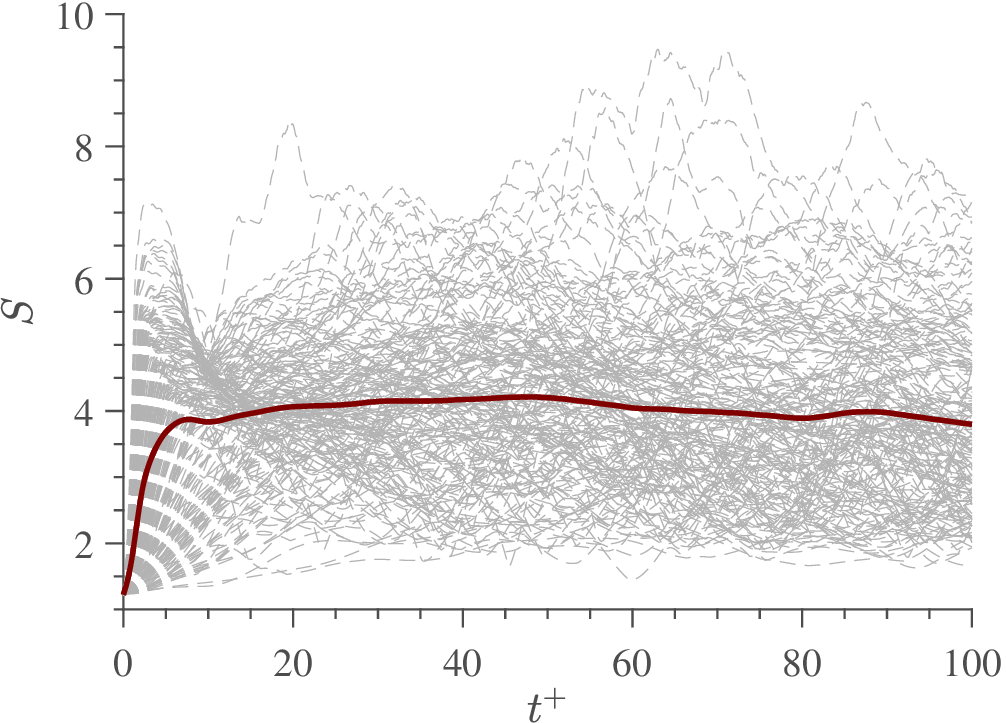}}\quad%
    \subfloat[ \label{sub:indivs180A}]  {\includegraphics[scale=0.4]{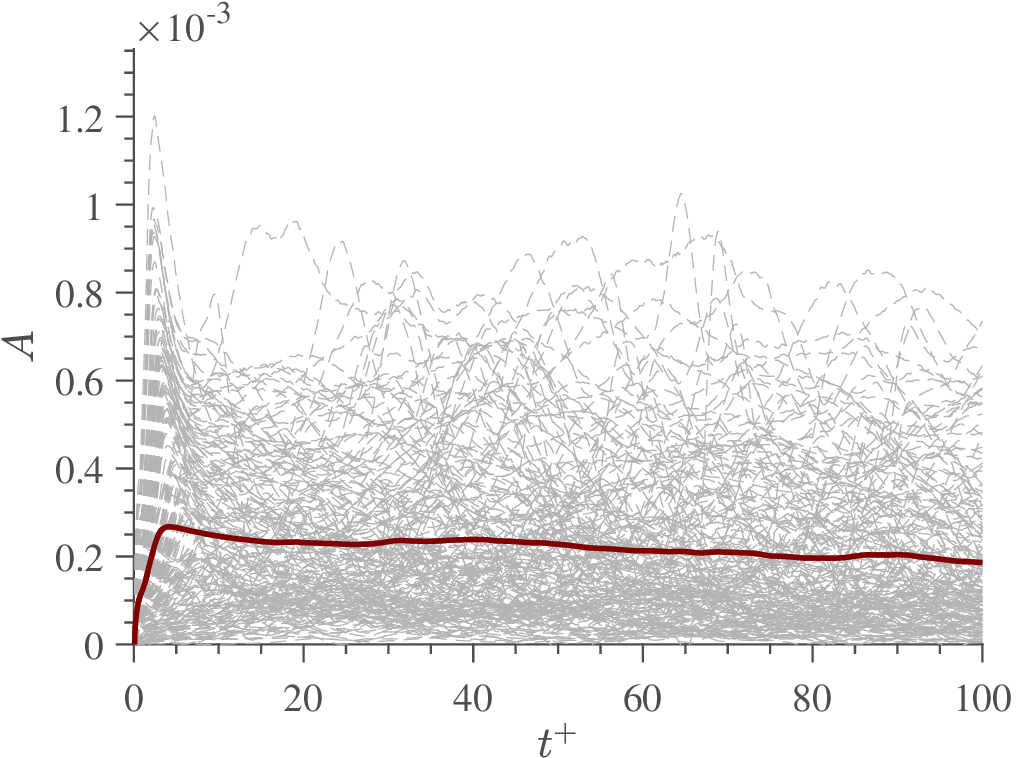}}\quad\\%
    \subfloat[ \label{sub:indivsL}]   {\includegraphics[scale=0.4]{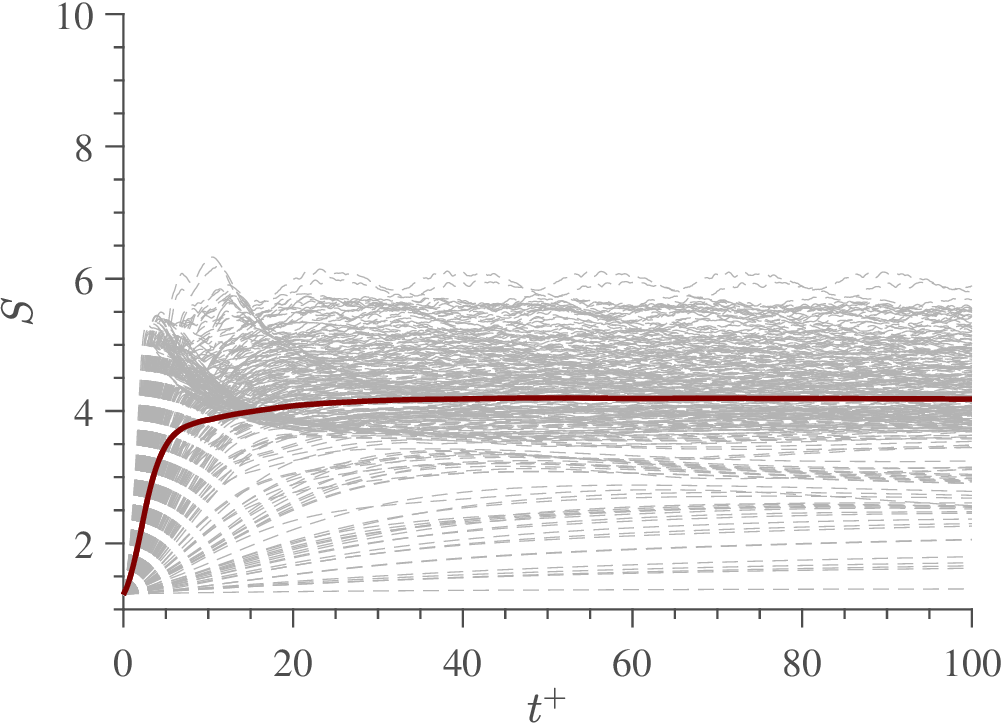}}\quad%
    \subfloat[ \label{sub:indivsLA}]  {\includegraphics[scale=0.4]{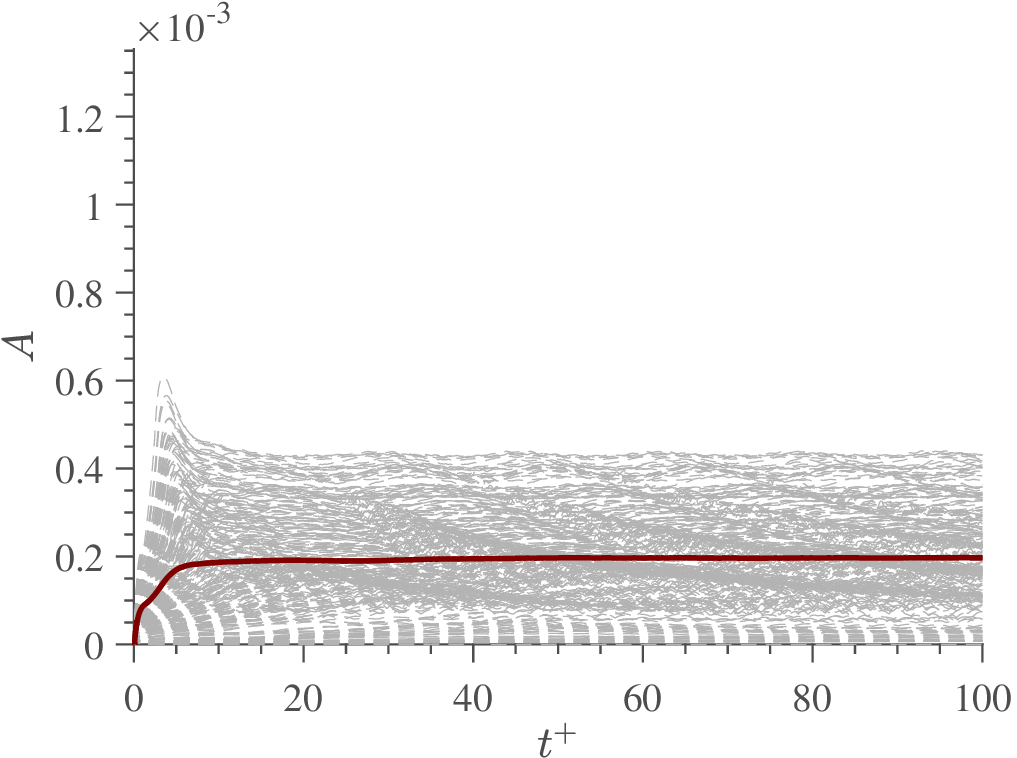}}\quad%
    \caption{A sample history of the deformation parameters $S$ (left column) and $A$ (right column) on a collection of cells, superimposed with the time-varying ensemble average of these individual cells $\langle S\rangle_c$ and $\langle A\rangle_c$, at a wall shear stress of $\tau_w = 30$ Pa. Results are plotted for the $Re_\tau =360$ (top row), $Re_\tau= 180$ (middle row), and laminar cases (bottom row).}
    \label{fig:indivs}
\end{figure}

Figure \ref{fig:indivs} displays samples of the history of $S$ and $A$ for all cells for the three cases described above: $Re_\tau=360$, $Re_\tau=180$, and the laminar case. These simulations are run at a wall shear stress of $\tau_w=30$ Pa. The time-varying ensemble average $\langle \bullet\rangle_c$ is superimposed on these plots.

\begin{figure}[H]
  \centering
  \subfloat[ \label{sub:avgs}] {\includegraphics[scale=0.38]{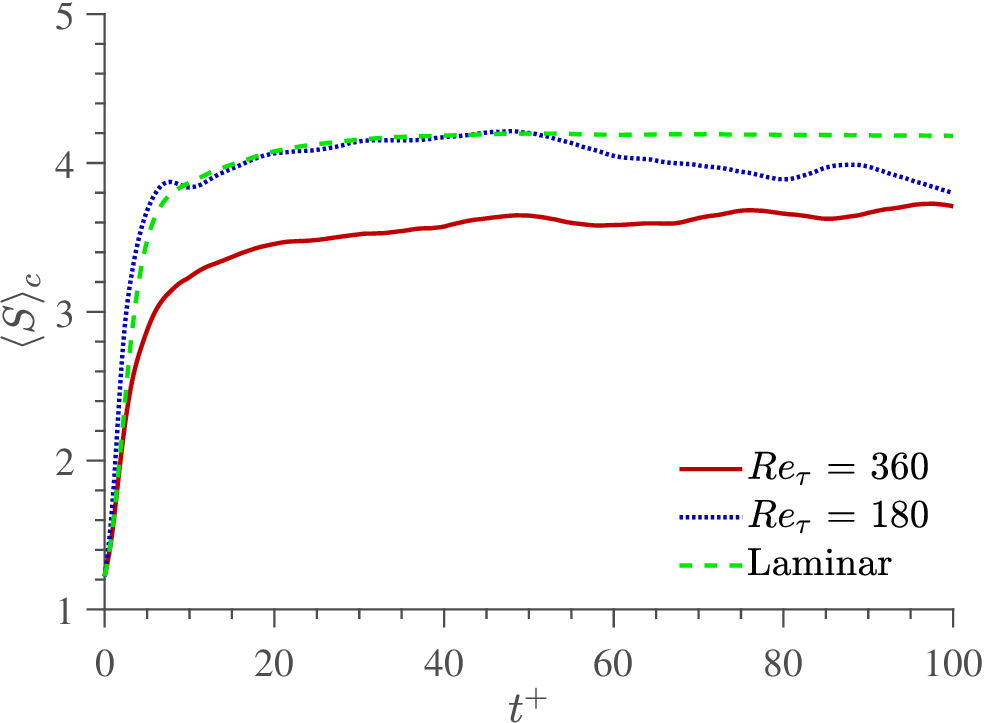}}\quad%
  \subfloat[ \label{sub:avgsA}] {\includegraphics[scale=0.38]{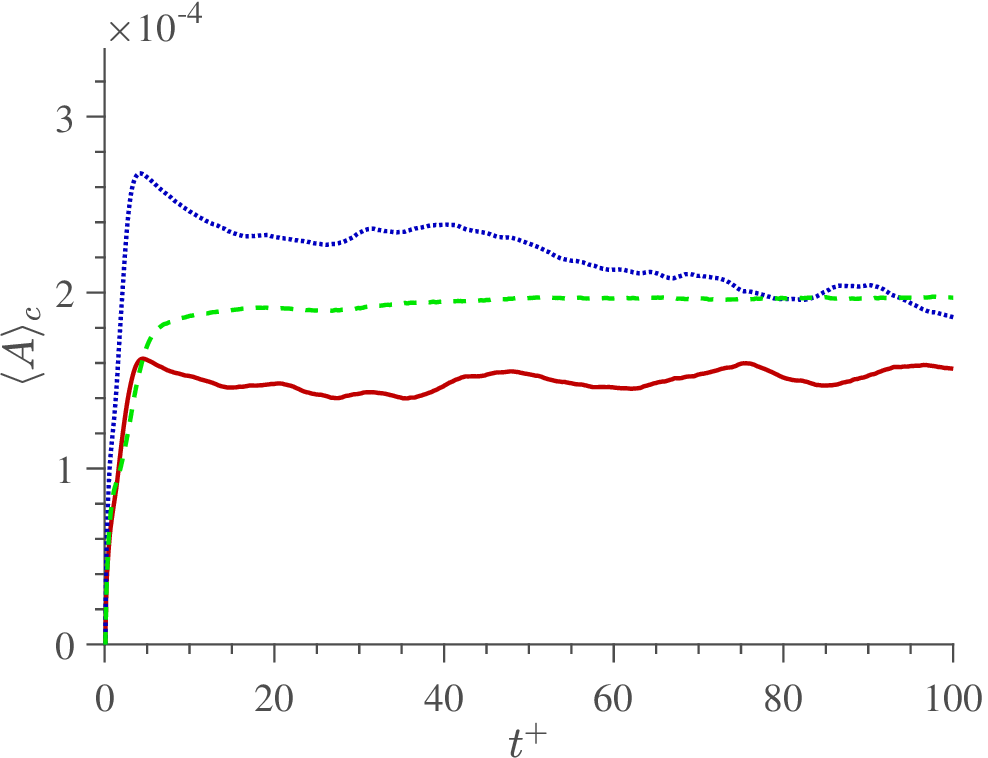}}\quad%
  \caption{The time-varying ensemble average $\langle S \rangle_c$ (left) and $\langle A \rangle_c$ (right) for the results shown in Fig. \ref{fig:indivs} at $\tau_w = 30$ Pa displayed on a single figure.}
  \label{fig:avgs}
\end{figure}

Figure \ref{fig:avgs} displays the time-varying ensemble averages in Fig. \ref{fig:indivs} on the same figure.

\begin{figure}[H]
    \centering
    \subfloat[ \label{sub:pdfstack360}]  {\includegraphics[scale=0.4]{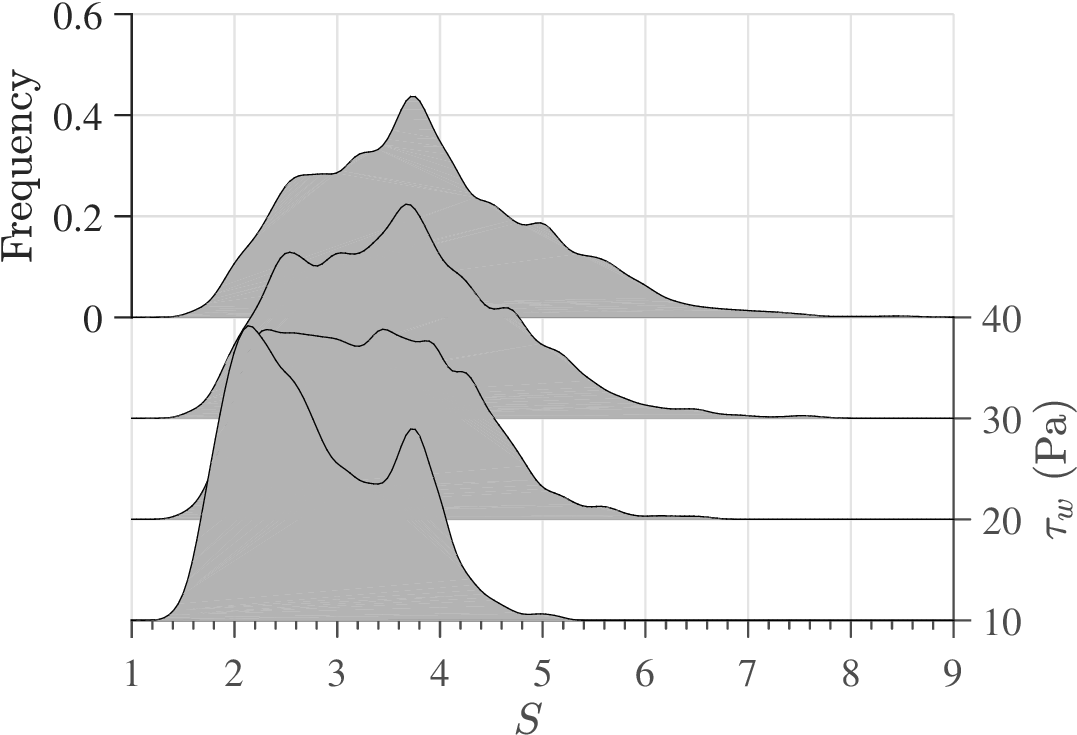}}\quad%
    \subfloat[ \label{sub:pdfstack360A}] {\includegraphics[scale=0.4]{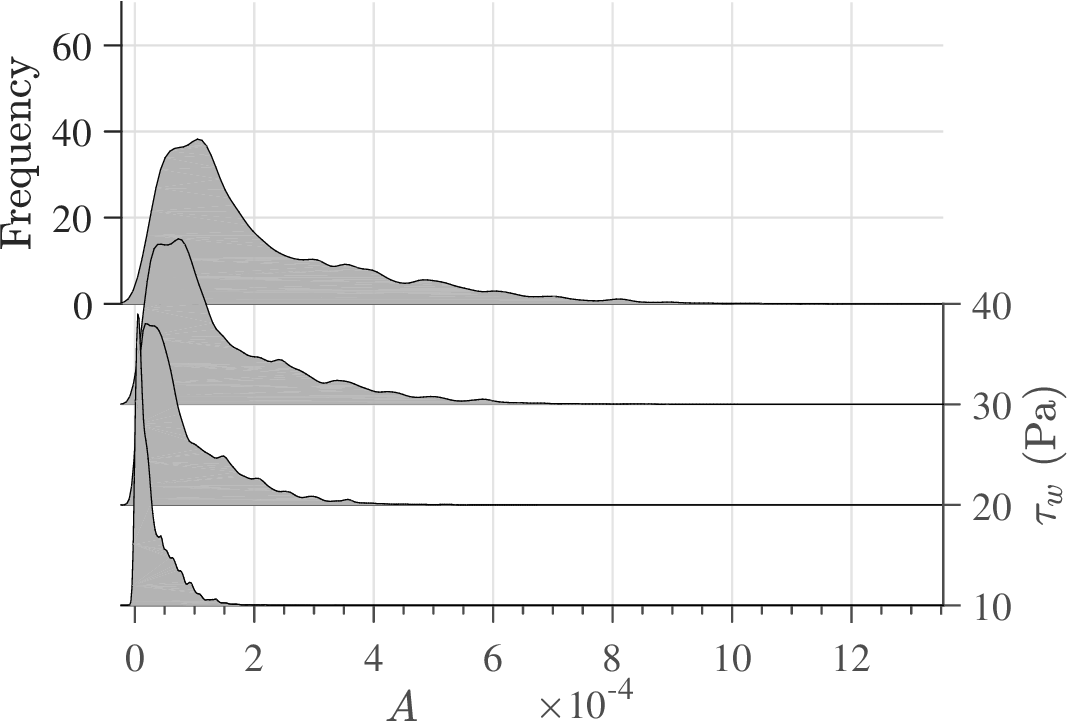}}\quad\\%
    \subfloat[ \label{sub:pdfstack180}]   {\includegraphics[scale=0.4]{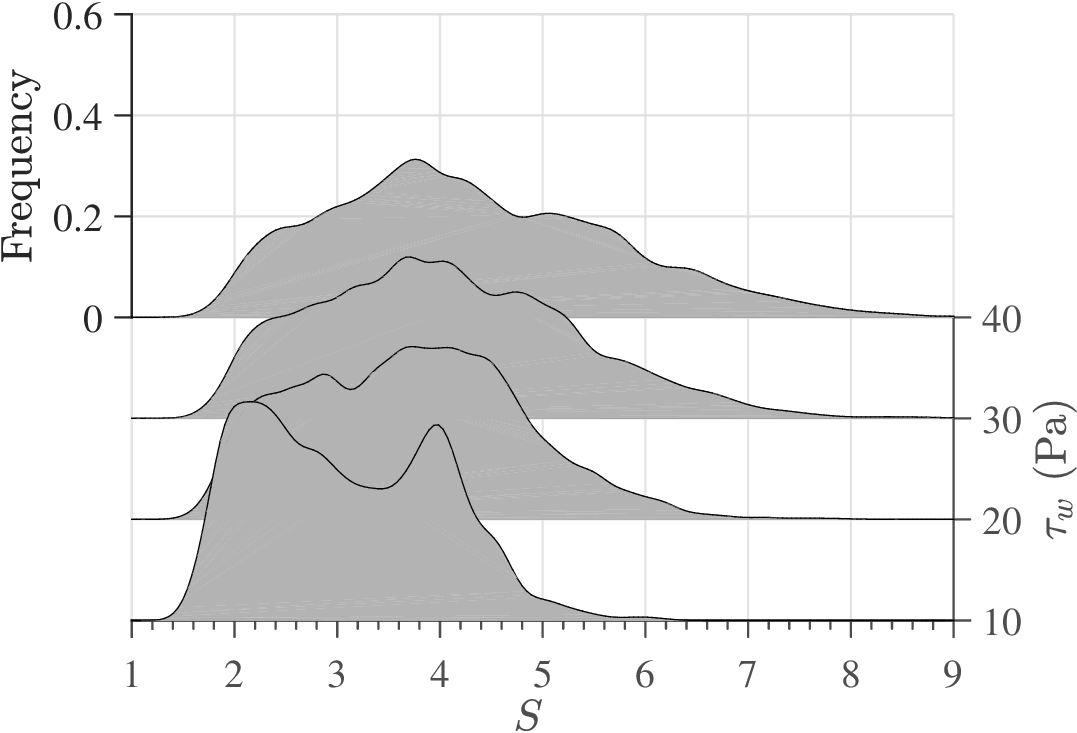}}\quad%
    \subfloat[ \label{sub:pdfstack180A}]  {\includegraphics[scale=0.4]{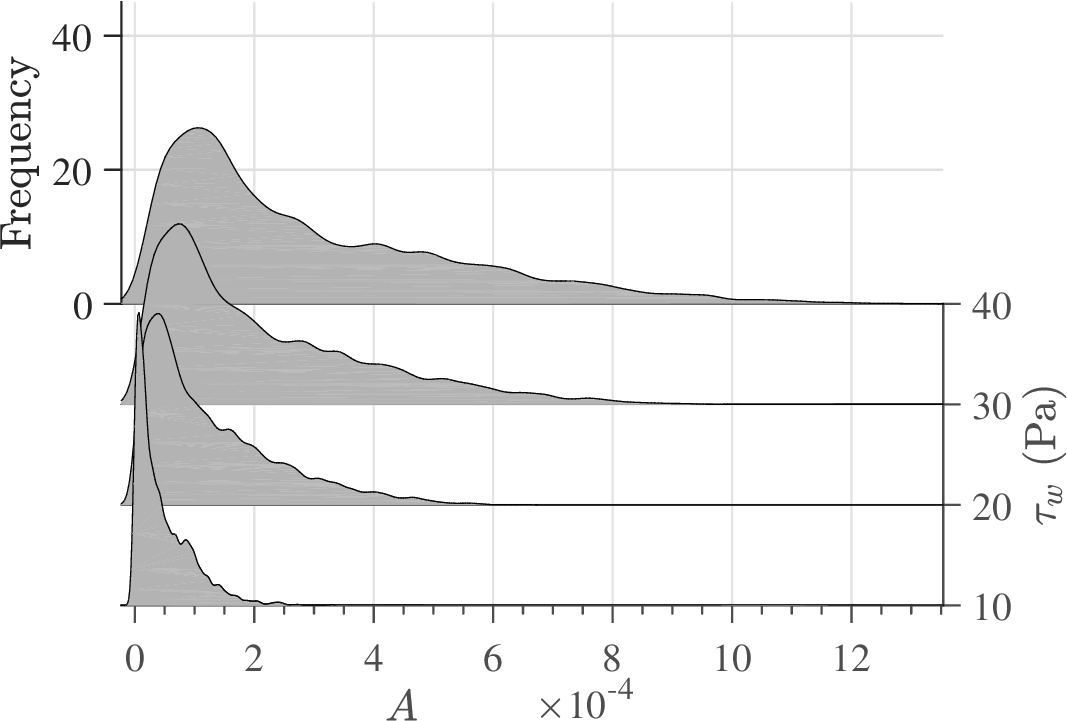}}\quad\\%
    \subfloat[ \label{sub:pdfstackL}]   {\includegraphics[scale=0.4]{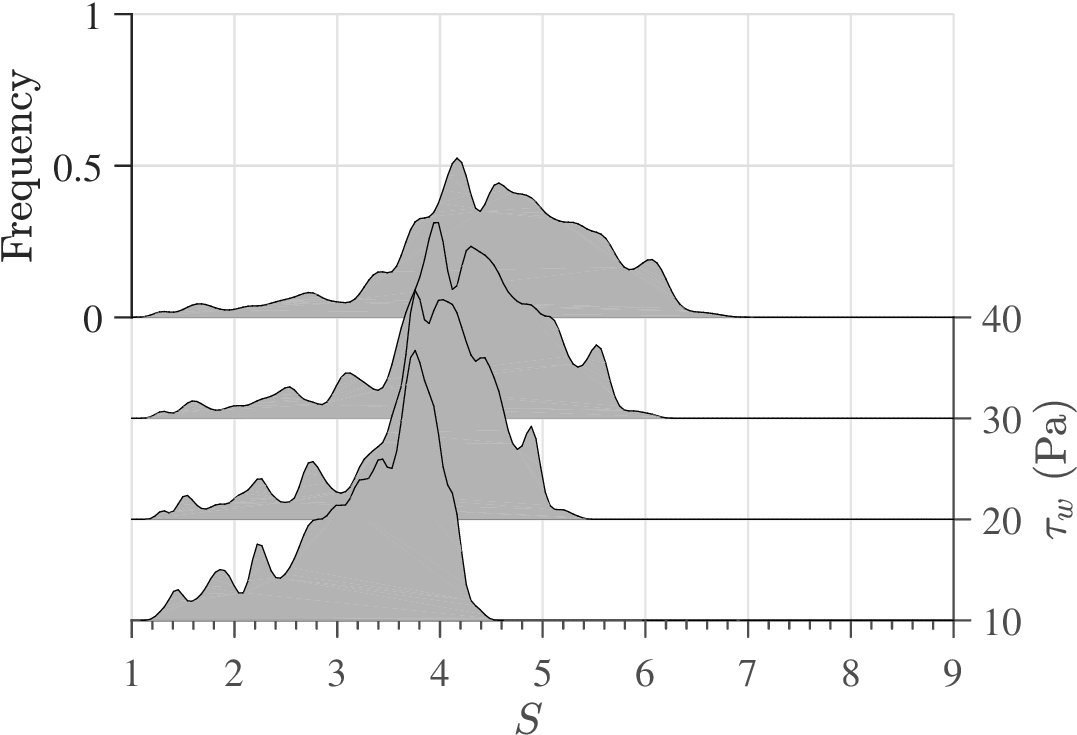}}\quad%
    \subfloat[ \label{sub:pdfstackLA}]  {\includegraphics[scale=0.4]{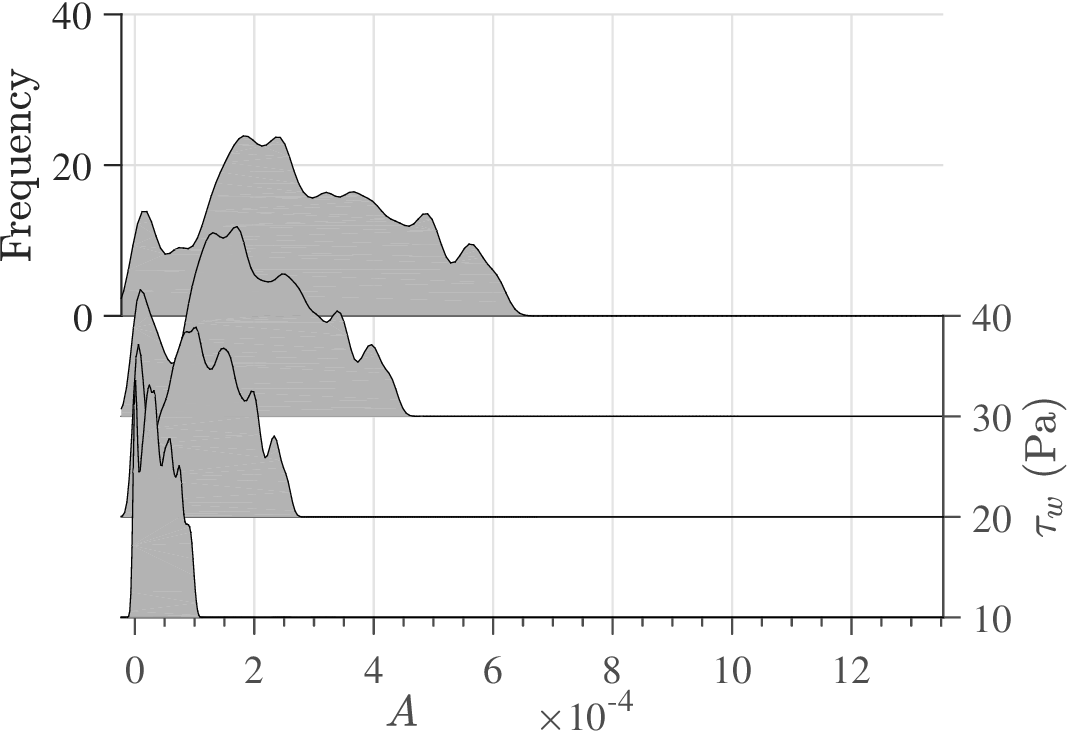}}\quad%
    \caption{PDFs of the parameters $S$ (left column) and $A$ (right column) as function of wall shear stress. Results are plotted for the $Re_\tau =360$ (top row), $Re_\tau= 180$ (middle row), and laminar cases (bottom row).}
    \label{fig:pdfstack}
\end{figure}

The above figures display a trend that is present across all values of the wall shear stress: The average values of the parameters $S$ and $A$ match at least somewhat well at the same value of $\tau_w$ across flow types, with some variation. However, the defining feature of the turbulent simulations relative to the laminar simulations is the greater content of cells that have particularly large deformation. In order to view this phenomenon from a more quantitative perspective, Fig. \ref{fig:pdfstack} displays PDFs for $S$ and $A$. These PDFs are constructed using the sum total of all individual time points from the individual cells. Since the RBCs enter the simulation undeformed from their biconcave resting shape, these PDFs are constructed based on the data after $t^+ >30$ to ensure they are independent from the initial conditions. In order to verify this criterion, a second PDF was constructed using half the window size to confirm that the features of the PDF did not change significantly. 

In Fig. \ref{sub:avgsA}, $\langle A\rangle_c$ appears to decrease with time somewhat for the $Re_\tau=180$ case. To ensure that this trend is temporary, we continued this specific simulation up to $t^+=400$. This calculation shows that $\langle A\rangle_c$ plateaus after the depicted range in Fig. \ref{sub:avgsA}, thus reaching a statistically stationary condition. Additionally, to ensure the reported results are not affected by a short integration time, we compared PDFs that were constructed over a time range of $30<t^+<100$ against those constructed for $65<t^+<100$. This comparison, and also an additional comparison made against data extracted from the extended simulation, showed that all PDFs are very similar to one another, particularly in the tails. Therefore, the integration period has a negligible effect on the results reported below. 

To obtain a measure of the RBC relaxation period to the background flow state, we employed $\langle S\rangle_c$ and $\langle A\rangle_c$ to construct a timescale for the RBCs. More specifically, a least-squares fit was performed with $\langle\bullet\rangle_c$ and the function
\begin{equation}\label{eq:fit}
    f(t^+; \tau^+)=\left(1-e^{-t^+/\tau^+}\right)\left(a_\infty -a_0 \right) +a_0,
\end{equation}
where $\tau^+$ is the associated time scale in wall units, $a_0$ is the initial value of the parameter and $a_\infty$ is the converged value of the parameter. This function is fitted with both $A$ and $S$. 

\begin{table}[H]
\centering
\caption{The time scale $\tau^+$ associated with the RBCs at all values of $\tau_w$ for the three Reynolds numbers of interest and the shear $S$ and area dilatation $A$ parameters. Results are obtained from a least squares fit to a curve expressed by Eq. \ref{eq:fit}.}
\label{tab:taus}
\begin{tabularx}{14cm}{c |X|X|X|X|X|X}
\multicolumn{7}{c}{$\tau^+$} \\
\hline
   \multicolumn{1}{c|}{} 
 & \multicolumn{2}{c|}{$Re_\tau=180$}  
 & \multicolumn{2}{c|}{$Re_\tau=360$}  
 & \multicolumn{2}{c}{Laminar}\\
 \hline
 $\tau_w$ (Pa) & $S$ & $A$ & $S$ & $A$& $S$ & $A$\\
 \hline
 10 & 1.9 & 0.55 & 2.6 & 0.42 & 2.9 & 1.2 \\
 20 & 2.1 & 0.93 & 3.0 & 0.88 & 3.3 & 2.4 \\
 30 & 2.3 & 1.1 & 3.3 & 1.1 & 3.5 & 2.6 \\
 40 & 2.6 & 1.2 & 3.6 & 1.2 & 3.8 & 2.6 \\
\end{tabularx}
\end{table}

The values reported in Table \ref{tab:taus} have a few implications. 
\begin{enumerate}
    \item They provide additional support for the fact that the simulations have been integrated for a sufficiently long time to allow the RBCs to adjust to changes in the flow. That is so since $\tau^+<10$ in all cases whereas computations are continued to $t^+=100$. 
    \item $\tau^+$ may serve as a relevant time scale for normalization of the residence time, which in conjunction with shear rate, is often taken as an input parameter to Eulerian hemolysis algorithms. The fact that $\tau^+ = O(1)$ for all cases studied here suggests that the dissipation (or Kolmogorov) time scale may be an appropriate choice for this purpose. 
    \item $\tau^+$ is generally shorter for the turbulent than the laminar case. This is particularly true for $A$ where $\tau^+$ is two to three times shorter in turbulent than laminar flow. One potential explanation for this gap can be obtained by considering a damped vibrating system, which settles quicker when subjected to an oscillatory external force than a uniform force. In the present case, turbulent flow subjects RBCs to a time-varying far-field boundary condition, thus permitting RBCs to come to an equilibrium state faster than if they were to be subjected to a uniform laminar flow. 
    \item A direct relationship between $\tau^+$ and $\tau_w$ is observed. This direct dependence is explained by the fact increasing $\tau_w$ reduces the flow time scale, which in turn increases the Capillary number, which can be conceptualized as the ratio of the membrane response time to a characteristic flow time. At lower values of $\tau_w$, i.e., between $\tau_w=10$ to 20 Pa, the membrane response time is almost a constant, hence producing a direct relationship between $\tau^+$ and $\tau_w$. At higher values of $\tau_w$, the membrane strain hardens. This reduces the time scale associated with the membrane, thereby, reducing the growth rate of $\tau^+$ with $\tau_w$. 
\end{enumerate}

Figure \ref{fig:pdfstack} displays a series of probability density functions for $S$ and $A$ for each value of Reynolds number as a function of the wall shear stress. We observe a shift rightward for each case as the wall shear stress is increased. The increase in content of high deformation in the turbulent cases is present in the relatively long tails of these cases. Due to the relatively large peaks at lower $\tau_w$, the PDFs of $A$ are scaled to be approximately the same height across values of $\tau_w$ in order to compare these distributions at different values of $\tau_w$ and prevent the front distributions from obscuring those in the back.

To compare these values quantitatively, Table \ref{tab:pcts} lists the percentages of the deformation values in Fig. \ref{fig:pdfstack} for the turbulent cases that are greater than the absolute maximum values in the corresponding laminar case.

\begin{table}[H]
\centering
\caption{The percentage of deformation values in the turbulent cases that are greater than the absolute maximum of the corresponding laminar case, calculated from the above PDFs.}
\label{tab:pcts}
\begin{tabularx}{10cm}{c |X|X|X|X}
\multicolumn{5}{c}{\% of ensembles above absolute maximum laminar value} \\
\hline
   \multicolumn{1}{c|}{} 
 & \multicolumn{2}{c|}{$Re_\tau=180$}  
 & \multicolumn{2}{c}{$Re_\tau=360$}\\
 \hline
 $\tau_w$ (Pa) & $S$ & $A$ & $S$ & $A$\\
 \hline
 10 &8.51 & 11.5 & 1.83 &3.34\\
 20 &6.49 & 13.6 & 1.54 &4.18\\
 30 &6.48 & 12.7 & 1.57 &4.01\\
 40 &5.25 & 11.6 & 1.15 &3.76\\
\end{tabularx}
\end{table}

\begin{figure}[H]
    \centering
    \subfloat[ \label{sub:compMean}]  {\includegraphics[scale=0.4]{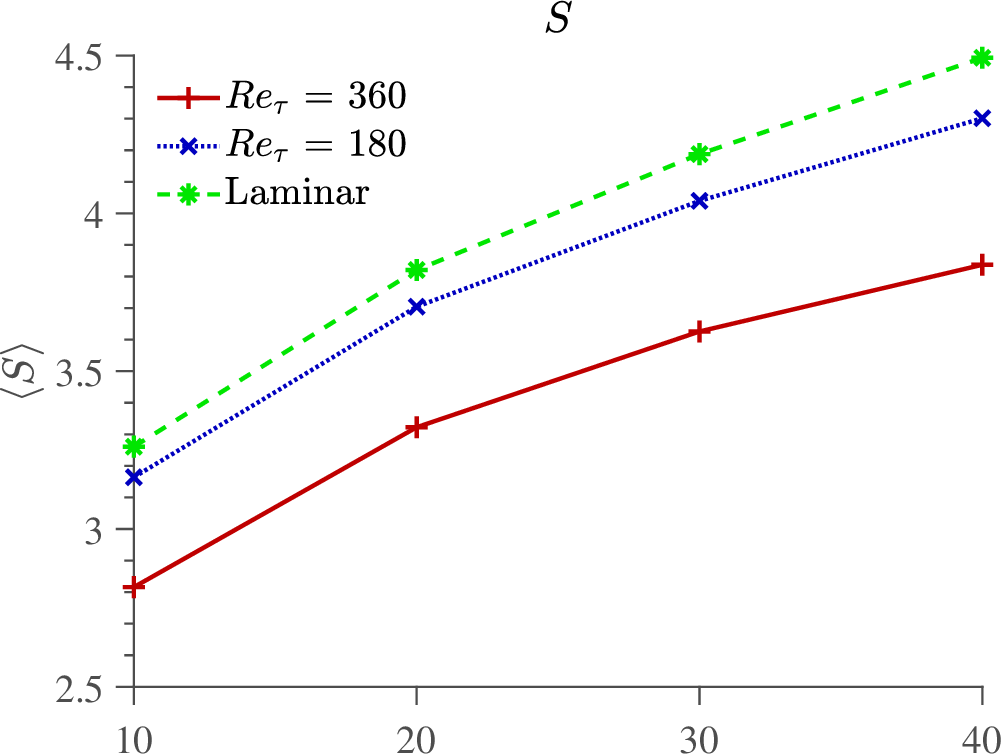}}\quad%
    \subfloat[ \label{sub:compMeanA}] {\includegraphics[scale=0.4]{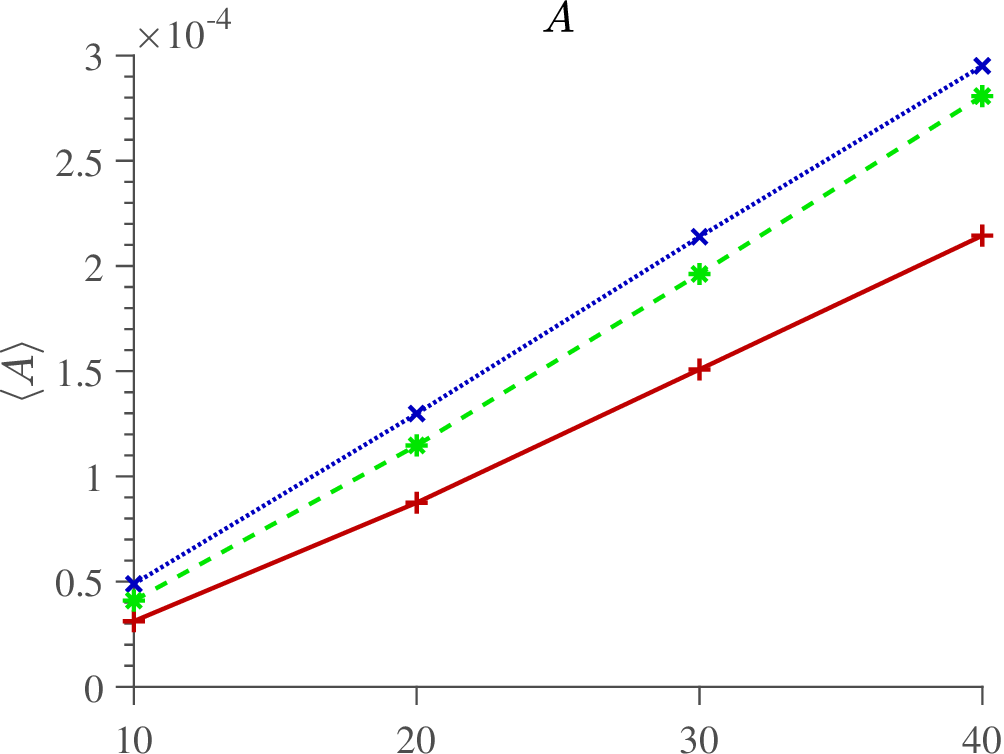}}\quad\\%
    \subfloat[ \label{sub:compStd}]   {\includegraphics[scale=0.4]{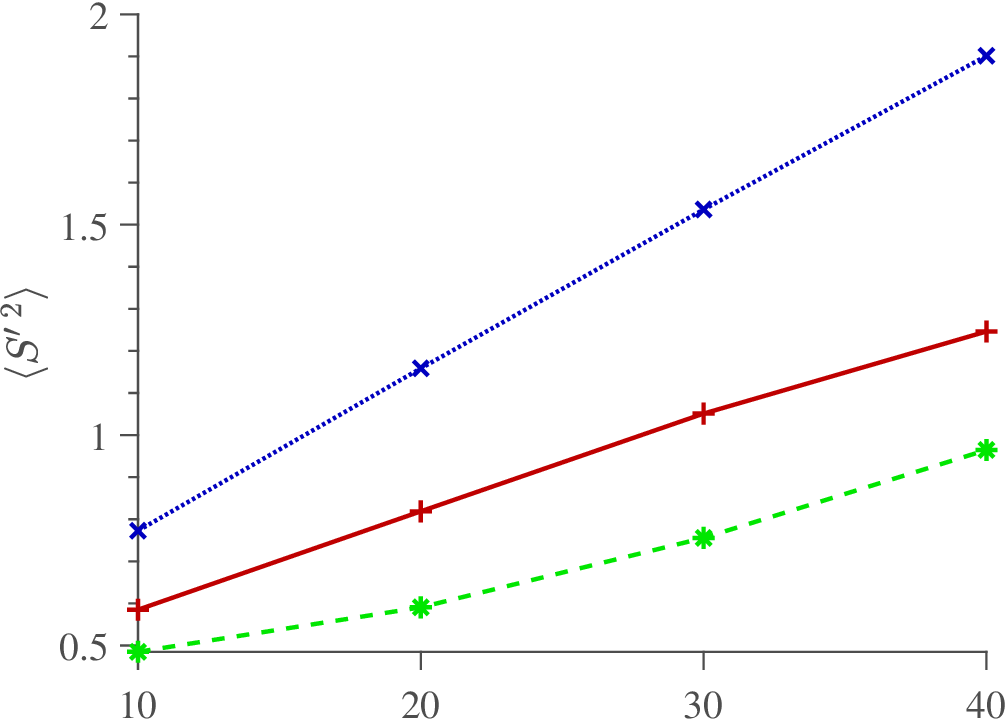}}\quad%
    \subfloat[ \label{sub:compStdA}]  {\includegraphics[scale=0.4]{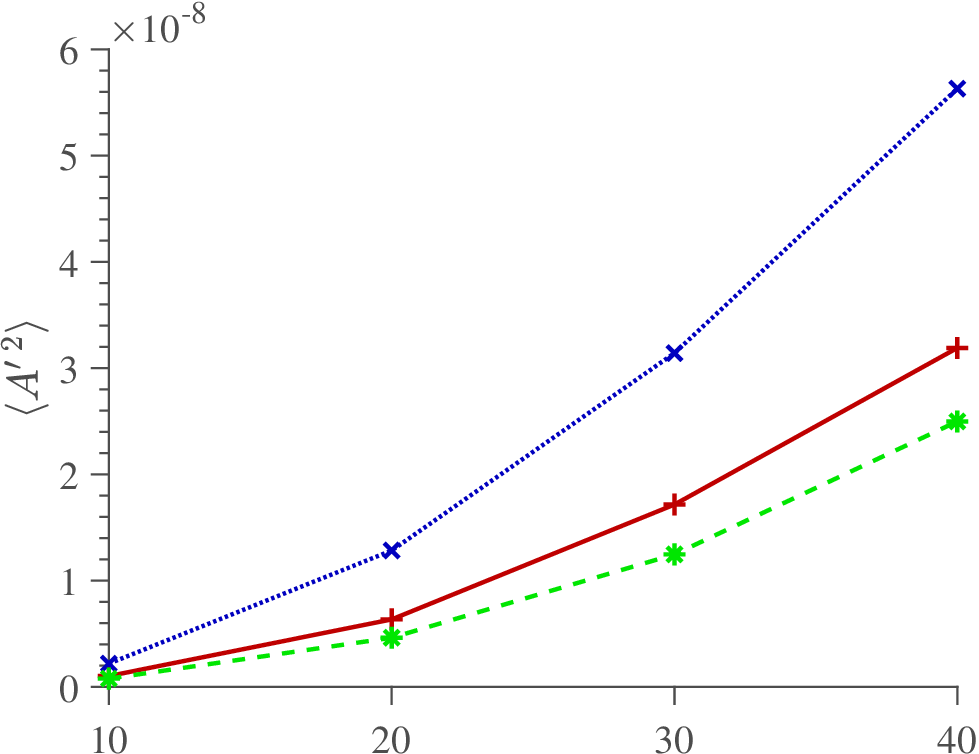}}\quad\\%
    \subfloat[ \label{sub:compMax}]   {\includegraphics[scale=0.4]{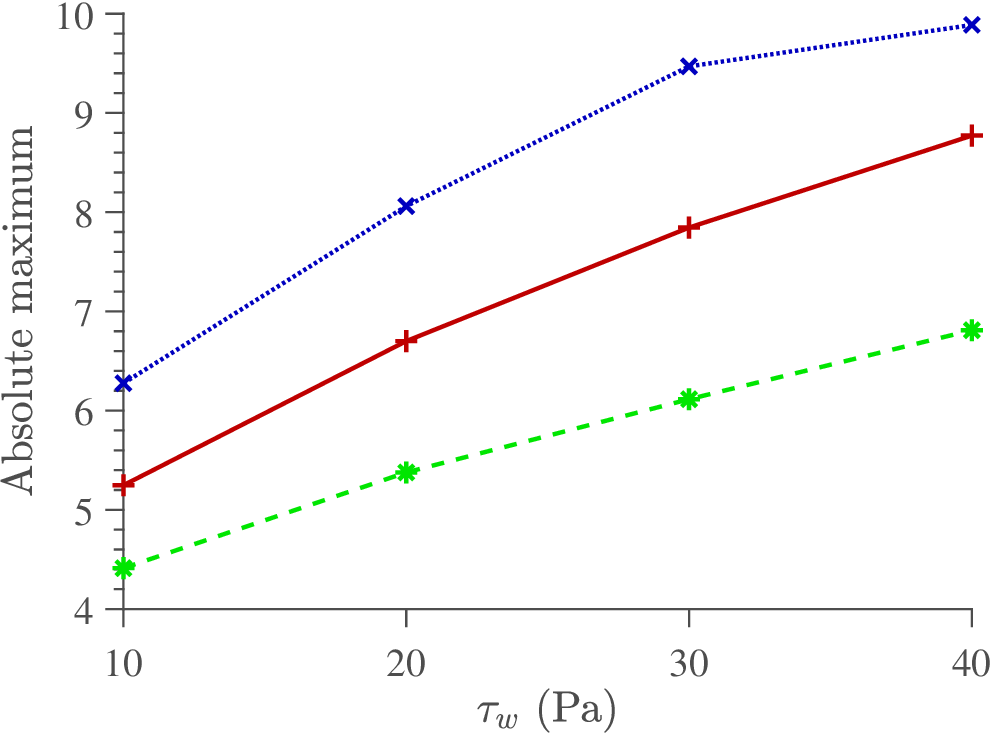}}\quad%
    \subfloat[ \label{sub:compMaxA}]  {\includegraphics[scale=0.4]{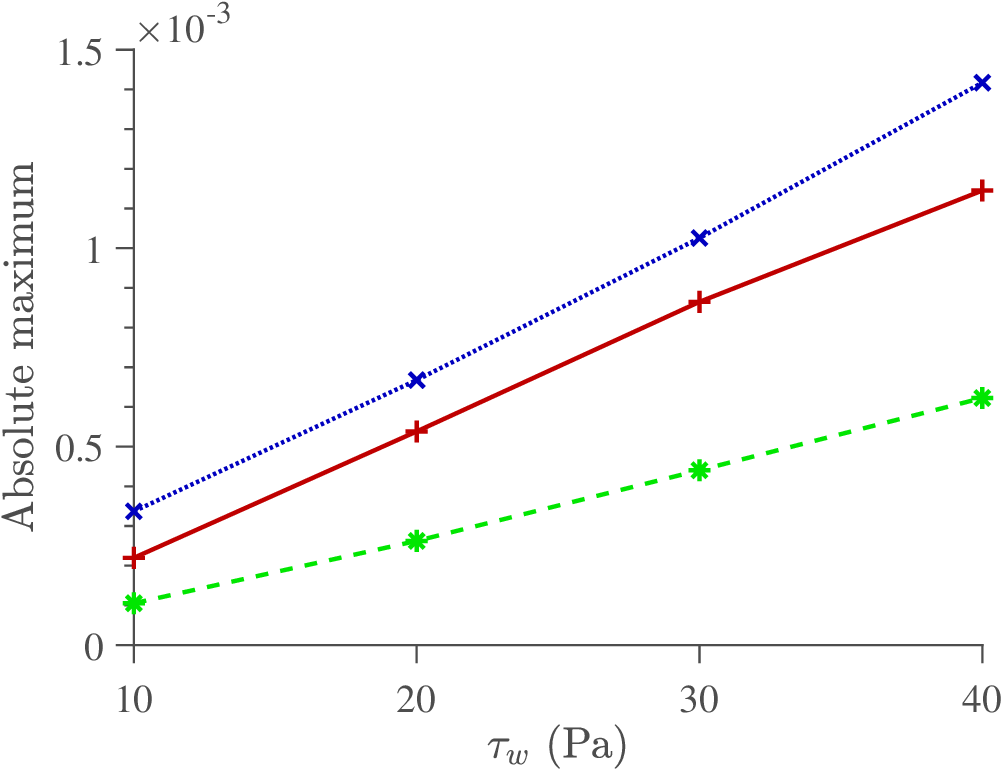}}\quad%
    \caption{Several features of $S$ (left column) and $A$ (right column) as a function of wall shear stress. The rows display, respectively, the ensemble average across all cells and time, their variance, and their absolute maximum value across all ensembles. Calculations are performed based on the data acquired at $t^+>30$.}
    \label{fig:comp}
\end{figure}

Finally, Fig. \ref{fig:comp} displays several features of $S$ and $A$ as a function of wall shear stress. This figure displays for these parameters the ensemble average across all time and RBCs, the ensemble average of the variance of the individual cells, and the absolute maximum value reached across all cells. The variance for a single cell is given as the average deviation from the mean, squared, described as $\langle\bullet'\,^2\rangle_t$. Note that the mean here is across all time and cells, such that $\bullet' = \bullet - \langle\bullet\rangle$. As before, calculations are only performed at $t^+>30$.

Empirical relationships for hemolysis typically rely on a bulk flow quantity (e.g., shear rate) to model hemolysis. Next, we investigate what will be relevant parameters from the flow that determine $S$ and $A$ in turbulence. For this purpose, the Pearson correlation coefficient is calculated between $S$ and $A$ and multiple features of the velocity gradient in what follows. In practice, bulk flow parameters are often extracted from flow solvers that do not resolve small dissipation scales in the flow. These RANS or LES-type solvers at best produce quantities that are averaged over time and space, thus filtering out the fine scale from the solution. To imitate this filtering process in our data, which contains those fine scales present in the DNS data, we sub-averaged quantities over a period of 15 in-wall units before treating them as independent points. 

The resulting correlation coefficients are reported in Table \ref{tab:corrco}. For bulk flow parameters we considered $\dot \gamma^+$ and the $Q$-criterion $Q=0.5(||\bl \Omega||^2-||\bl E||^2)$. Here, $E_{ij} = 0.5(\nabla_ju_i + \nabla_iu_j)$ and $\Omega_{ij} = 0.5(\nabla_ju_i - \nabla_iu_j)$ are the symmetric and antisymmetric components of the velocity gradient tensor. Correlation coefficients are also calculated against $\|\bl E\|$ and $\|\bl \Omega\|$. Note that $\dot \gamma^+ = \|\nabla \bl u\|$ and $\|\bl E\|^2+\|\bl \Omega\|^2 = \|\nabla \bl u\|^2$. These calculations are performed at different values of $\tau_w$. The range displayed in Table \ref{tab:corrco} is selected to encompass the correlation coefficients at all values of $\tau_w$. 

\begin{table}[H]
\centering
\caption{The ranges of Pearson correlation coefficients between RBC deformation and bulk flow parameters across values of $\tau_w$.}
\label{tab:corrco}
\begin{tabularx}{12cm}{c |X|X|X|X}
\multicolumn{5}{c}{Pearson correlation coefficient $R$} \\
\hline
   \multicolumn{1}{c|}{} 
 & \multicolumn{2}{c|}{$Re_\tau=180$}  
 & \multicolumn{2}{c}{$Re_\tau=360$}\\
 \hline
  & $S$ & $A$ & $S$ & $A$\\
 \hline
 $S$        & $+1.00\pm0.00$  & $+0.92\pm0.02$ & $+1.00\pm0.00$ & $+0.90\pm0.02$ \\
 $A$        & $+0.92\pm 0.02$ & $+1.00\pm0.00$ & $+0.90\pm0.02$ & $+1.00\pm0.00$ \\
 $\dot \gamma^+$ & $+0.87\pm0.02$  & $+0.97\pm0.01$ & $+0.82\pm0.02$ & $+0.95\pm0.01$ \\
 $Q$        & $-0.24\pm0.01$  & $-0.27\pm0.01$ & $-0.24\pm0.02$ & $-0.26\pm0.01$ \\
 $\|\bl E\|$        & $+0.85\pm0.02$  & $+0.95\pm0.01$ & $+0.80\pm0.02$ & $+0.93\pm0.02$ \\
 $\|\bl \Omega\|$   & $+0.88\pm0.02$  & $+0.98\pm0.01$ & $+0.83\pm0.01$ & $+0.97\pm0.02$ \\
\end{tabularx}
\end{table}
The results shown in Table \ref{tab:corrco} indicate the change in $\tau_w$ has minimal effect on $R$ since the range is less than $\pm 0.02$ for all cases. A strong correlation between RBC deformation with $\dot \gamma^+$, $\|\bl E\|$, and $\|\bl \Omega\|$ is also observed. The results also indicate a poor correlation of $S$ and $A$ with $Q$. We must note that the majority of ensembles fall near $Q\approx 0$, including large deformation incidents, suggesting that the regions where the the most intense bursts in velocity gradient occur are primarily shear flow.

\begin{figure}[H]
    \centering
    \subfloat[ \label{sub:reg360S}]   {\includegraphics[scale=0.43]{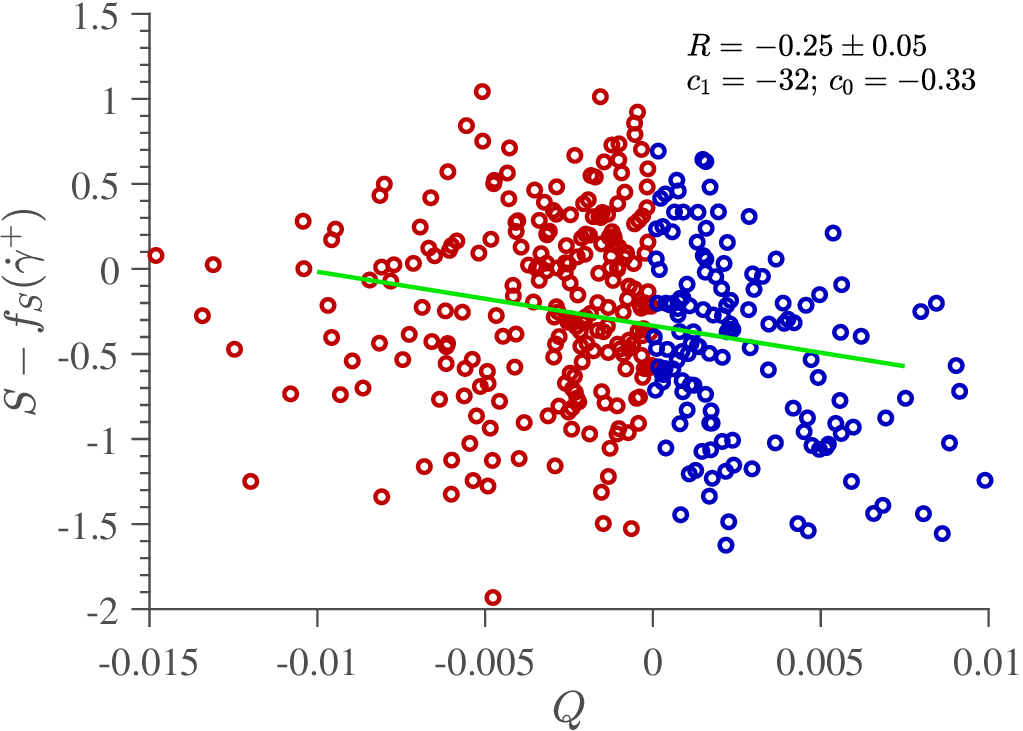}}\quad%
    \subfloat[ \label{sub:reg360A}]  {\includegraphics[scale=0.43]{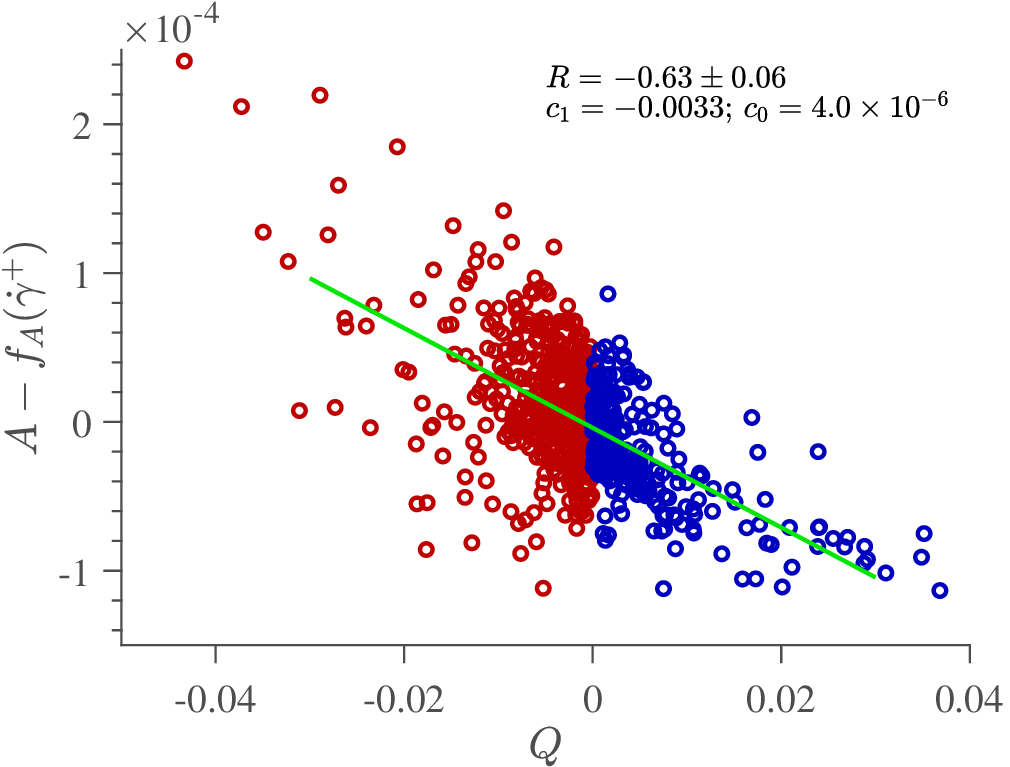}}\quad\\%
    \subfloat[ \label{sub:reg180S}]  {\includegraphics[scale=0.43]{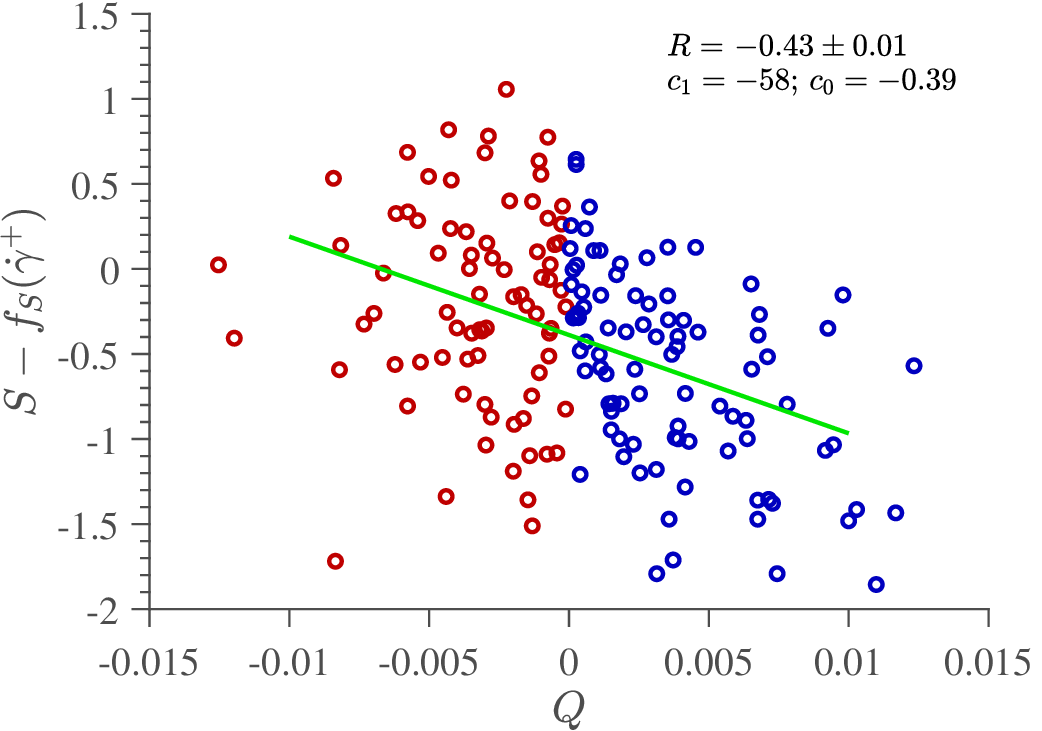}}\quad%
    \subfloat[ \label{sub:reg180A}] {\includegraphics[scale=0.43]{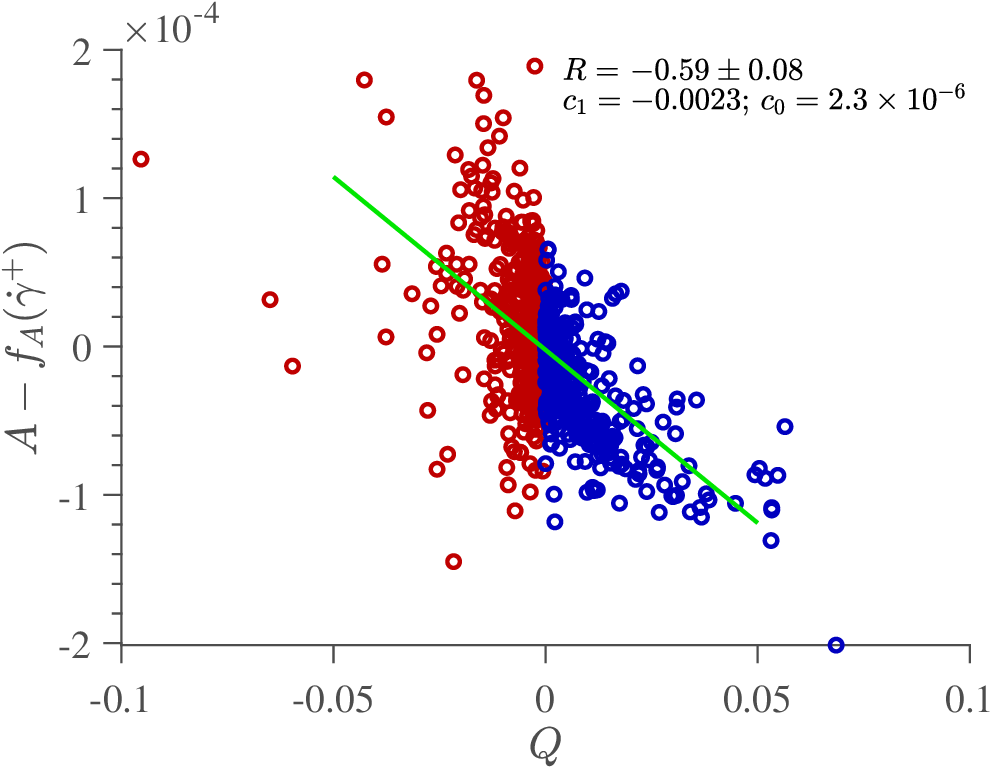}}\quad\\%
    \caption{Scatter plots of the adjusted values of $A$ and $S$ as a function of $Q$. Plots are for $Re_\tau=360$ (top) and $Re_\tau=180$ (bottom). The linear fit is superimposed on the data, with the constants of this line in the upper right. The Pearson correlation coefficient $R$ here refers to the range over all values of $\tau_w$ tested. Note that positive and negative $Q$ are differentiated here, as these values correspond to regions in the flow dominated by rotation and extension, respectively.}
    \label{fig:Regs}
\end{figure}

The laminar flow considered here exposes RBCs to pure shear where $\dot \gamma^+\ne 0$ and $Q=0$. To better understand the effect of incidents with $Q\ne 0$ on cell deformation, we must rely on turbulence data. Since turbulence data is also dominated by shear incidents, doing so requires adjusting $S$ and $A$  so that the effects of $\dot\gamma^+$ are removed from $S$ and $A$. This way, we can remove the first-order effects of $\dot \gamma^+$ from the $S$ or $A$ signals so that the secondary effect of $Q$ can be teased out. To establish the first-order baseline, we can rely on the laminar simulations data where $Q=0$, allowing us to measure the effect of $\dot \gamma^+\ne 0$ on $S$ and $A$. The full process is as follows. First, a linear fit with the laminar simulations was constructed for $S$ and $A$ with $\dot\gamma^+$, such that, for $A$, $f_A(\dot\gamma^+) = c_1\dot\gamma^+ + c_0$. For $A$, this range was taken over the entire range of $\dot\gamma^+$ available for the laminar flow. For $S$, it was found that a good fit was obtained for $f_S(\dot\gamma^+)$ at approximately $\dot\gamma^+<0.2$, with an average value of $R=0.91$. Thus, only this range was used for constructing the linear fit. Second, the adjusted value $A-f_A(\dot\gamma^+)$ is computed. This adjusted value represents the new signal which ``excludes'' the effect of $\dot \gamma^+$. Thirdly, the correlation between this adjusted $A-f_A(\dot \gamma^+)$ and $Q$ is computed. The result of these calculations is displayed in Fig. \ref{fig:Regs}, where the individual ensembles are shown as a scatter plot. This plot is displayed for $\tau_w = 30$ Pa. However, the correlation coefficients are displayed for the range of $\tau_w$. In this plot, the positive and negative values are displayed in separate colors, as these values correspond to incidents of rotational and extensional flows, respectively.

A few observations can be made based on the results shown in Fig. \ref{fig:Regs}. 
First, in all cases, the adjustment made above improved the correlation against $Q$. This is particularly true in the case of $A$ where $R$ improves from approximately $-0.26$ to around $-0.6$. 
Second, $S$ and $A$ are negatively correlated against $Q$. That is true regardless of whether those parameters are adjusted or not.  
Third, the adjusted $S$ has a worse correlation than the adjusted $A$. This is despite the fact that the ensembles associated with high shear ($\dot \gamma^+>0.2$) were excluded for $S$. This worse correlation, which also appears in Table \ref{tab:corrco}, can be explained by more complex behavior of $S$ relative to $A$. This complex behavior of $S$ is not exclusive to turbulence and also appears in the laminar case under consideration, where it produced nonlinear behavior versus $\dot \gamma^+$ that prompted us to exclude data at $\dot \gamma^+>0.2$ in performing this secondary analysis. 

\FloatBarrier

% --------------------
\section{Discussion} \label{sec:discussion}
While Eulerian hemolysis algorithms typically take both the residence time and fluid stress as predictors of hemolysis severity \cite{giersiepen_estimation_1990, grigioni_power-law_2004}, many cell-resolved hemolysis algorithms are formulated in terms of the level of strain experienced by the cell \cite{nikfar_multiscale_2020, xu_cell-scale_2023, sohrabi_cellular_2017}, where cells do not experience damage until they reach a certain strain. Accumulated damage is typically a time-dependent function of the strain once damage begins to occur, however. We quantified strain experienced by the cell through $S$ and $A$, the PDFs of which were shown in Fig. \ref{fig:pdfstack}. The peaks of these PDFs occur at fairly similar values of $S$ and $A$, with the laminar cases typically peaking later than the turbulent cases. However, one recurrent feature of the turbulent cases relative to the laminar case is the fact that the turbulent cases have a longer tail that extends into higher deformation values, whereas the laminar case has a more concentrated PDF. Under many models of cell-resolved hemolysis, the RBCs will reach the threshold strain and begin to undergo damage in the turbulent cases faster than in the laminar case. In fact, if this threshold were, for example, at the maximum strain experienced in the laminar case, damage would be occurring in 14\% of the cells on average in the turbulent case when examining the worst-case scenario of Table \ref{tab:pcts}. These differences are partially described by the low variance in the laminar case relative to the turbulent cases, as seen in Figs. \ref{sub:compStd} and \ref{sub:compStdA}. While the mean values of the parameters in the turbulent case are generally less than or comparable to the laminar case, the variances are higher in the turbulent cases, leading to periods of elevated deformation and greater maximum stretch as seen in Figures \ref{sub:compMax} and \ref{sub:compMaxA}. As described above, these periods of relatively large strain are likely to have an outsized effect on RBC damage relative to the periods of lower, more constant strain that the cells undergo in the laminar flow case.

One interesting feature of the turbulent cases that persists across values of $\tau_w$ is the fact that the $Re_\tau =180$ case consistently produces larger strains than the $Re_\tau = 360$ case, both in terms of peak strain as well as in terms of the tail thickness. This is also shown in Fig. \ref{fig:comp}, in which the $Re_\tau=180$ case causes more deformation than the $Re_\tau=360$ case across all values of $\tau_w$ and in all measures. This is somewhat counterintuitive: one might expect that with increasing turbulence intensity, there would be increasing damage. While the relative spikes in velocity gradient would be expected to be higher in the case where the turbulence intensity is higher, in order to obtain the same values of $\tau_w$, other parameters would need to be adjusted in an experiment that could potentially impede damage to the RBC. For example, keeping other variables such as channel width and bulk channel velocity constant, this higher turbulence intensity could be achieved by lowering the viscosity, which would cause a resultant drop in shear stress experienced by the cells in the fluid. So while the turbulence intensity will cause larger spikes in the magnitude of the velocity gradient, this relationship has a less straightforward translation to the strain experienced by the RBC. Experimentally, increasing the Reynolds number is often associated with an increase in flow rate, and thus an increase in wall shear stress; however, the fact that the wall shear stress is held constant between simulations at different Reynolds numbers produces some slightly counter-intuitive results. What appears to be important when comparing laminar to turbulent flow, however, is the fact that the latter contains rare bursts of high-velocity gradient that cause relatively large strains in the cells that are absent in the former. This relationship is present in Fig. \ref{fig:inputs}, in which the average cell experiences a smaller normalized value of the velocity gradient at higher Reynolds flow on average.

As has been mentioned previously \cite{quinlan_mechanical_2014}, the wall shear stress may not necessarily provide a direct analogy between turbulent and laminar flows. As seen in Fig. \ref{fig:inputs}, the average value of the stress is significantly smaller in the case of the turbulent flows than the laminar flow, and this average value decreases with increasing $Re_\tau$. However, one would expect this difference to result in an increase in deformation in the laminar case relative to the turbulent case, all else equal. Despite this, the turbulent simulations still produce more deformation than the laminar simulations when considering maximum deformation.

Also of note in the results is the fact that the cells undergo relatively smaller changes as the wall shear stress is increased, particularly for $S$. Doubling the wall shear stress did not double RBC deformation. This is likely a result of the strain-hardening nature of RBCs, in which the cells' resistance to deformation increases the more they are deformed. As a result, a small increase in strain causes a large increase in stress at large deformation, and a small increase in stress causes a large increase in strain at low deformation.

The large values of the correlation coefficient with respect to $\dot\gamma^+$, particularly for $A$, suggest that the largest deformation in the flows studied here occurs in strong shear. This observation justifies the use of $\dot\gamma^+$ as a primary parameter in hemolysis models that rely on bulk flow parameters for similar types of flows as those studied here. Additionally, a moderate correlation emerged between cell deformation metrics and $Q$ after excluding the effects of $\dot\gamma^+$. This observation suggests that taking in $Q$ as an independent input in those hemolysis models can improve their predictability of cell deformation in turbulent flows. That is particularly true if such models rely on $A$ rather than $S$ for its better secondary correlation that applied to a broader range of conditions. Furthermore, the fact that the secondary correlation was negative suggests that a deviation from shear flow toward extensional flow will cause more deformation than a deviation toward a rotational flow. We reiterate that $Q$ is of secondary importance in modeling $A$ or $S$ in turbulence as most of the large deformation events here are associated with shear events in the turbulent simulations. Given that in the laminar case considered here $Q=0$, the differences between the laminar and turbulent cases can be attributed to the spikes in velocity gradient and also the secondary effects discussed above.

\FloatBarrier
% --------------------
\subsection{Future work}\label{sec:future}
Multiple facets of turbulent interactions with RBCs still remain to be studied. For example, the nature of channel flow (and indeed the pipe flow studied in experiments by Kameneva et al \cite{kameneva_effects_2004}) means that the cells in the laminar flow are locally exposed only to a constant shear flow. This could potentially be responsible for some differences between the laminar and turbulent simulations, as the RBCs in the turbulent simulations are exposed to a range of different velocity fields. It would be valuable to expose the cells to different stress configurations in the laminar flow case, e.g., extensional flow, and compare those results to analogous turbulent cases.

Additionally, simulations of RBCs under large strain rates for extended periods of time are necessary to study hemolysis from a cell-resolved perspective. Much of the study of RBCs using mesoscopic methods has centered on Capillary numbers too small to create cell damage. Thus, extending simulations to higher values of shear rate is a next step in the study of cell-resolved hemolysis.

Finally, a model linking these deformation parameters to RBC damage should be implemented. While studies such as this one allow comparison of RBC deformation in a relative sense between different cases (e.g., comparison of similar flows at equivalent or slightly different wall shear stress), implementing a damage model will allow geometries to be evaluated in an absolute sense in terms of how much damage they actually produce. Such models will likely need to be empirical, but the framework used in the current work for obtaining data on the deformation of a large number of cells is invaluable for indirect measurement of cell deformation from experiments in complex flow conditions.

% --------------------
\section{Conclusions}\label{sec:conclusions}
We investigated the effect that turbulence had on RBCs traversing a channel flow. This comparison was facilitated using a boundary integral solver, which took in the Lagrangian velocity gradient from the channel flow simulation as an input. Inspired by the experiments by Kameneva et al \cite{kameneva_effects_2004}, who showed that turbulence can cause an increase in hemolysis when compared to laminar flows at the same wall shear stress, laminar and turbulent simulations were compared at equivalent wall shear stress. Additionally, the turbulent simulations were run at two values of the friction Reynolds number, $Re_\tau = 180$ and $360$. Although simulations were run at lower values of the wall shear stress than the experiments, the experiments were corroborated by the simulations in some key ways. First, the cells experienced greater maximum deformation in the turbulent than in the laminar simulations. On average, however, the red blood cells did not deform as much as in the laminar flow. On the other hand, the PDFs of the deformation in the turbulent simulations still had a wider tail than the laminar simulation, indicating a greater frequency of extreme events. Interestingly, the $Re_\tau = 360$ simulations did not produce more deformation than the $Re_\tau = 180$ simulations, indicating that turbulence intensity does not equate to greater cell damage when all other parameters are held equal. Filtered results extracted from turbulence cases showed that while $\dot\gamma^+$ is the primary parameter that correlates well with the cell deformation in this type of flow, one must also consider the $Q$-criterion as a secondary parameter to build a refined model of cell deformation from the bulk flow parameters in turbulence.

% --------------------
\section*{Acknowledgements}
Research reported in this publication was supported by the National Heart, Lung, and Blood Institute of the National Institutes of Health under award number R01HL089456-10.

%--------------------
% REFERENCES SECTION
\color{black}
\medskip
\bibliographystyle{elsarticle-num}
\bibliography{references} 

\newpage

\end{document}